\documentclass[twocolumn,showpacs,amsmath,pre]{revtex4}
\usepackage{graphicx}
\usepackage{amsmath}
\usepackage{amssymb}
\usepackage{natbib}
\usepackage{subfigure}

\begin{document}

\title{Constraints and vibrations in static packings of ellipsoidal particles}

\author{Carl F. Schreck$^{1}$}
\author{Mitch Mailman$^{2}$}
\author{Bulbul Chakraborty$^{3}$} 
\author{Corey S. O'Hern$^{4,1}$} 

\affiliation{$^{1}$Department of Physics, Yale University, New Haven,
  Connecticut 06520-8120, USA}

\affiliation{$^{2}$Department of Physics, University of Maryland,
  College Park, Maryland 20742, USA}

\affiliation{$^{3}$Martin Fisher School of Physics, Brandeis University, Mail Stop 057,
Waltham, MA 02454-9110, USA}

\affiliation{$^{4}$Department of Mechanical Engineering \& Materials Science, Yale University, New
  Haven, Connecticut 06520-8260, USA}

\begin{abstract}
We numerically investigate the mechanical properties of static
packings of ellipsoidal particles in 2D and 3D over a range of aspect
ratio and compression $\Delta \phi$. While amorphous packings of
spherical particles at jamming onset ($\Delta \phi=0$) are isostatic
and possess the minimum contact number $z_{\rm iso}$ required for them
to be collectively jammed, amorphous packings of ellipsoidal particles
generally possess fewer contacts than expected for collective jamming
($z < z_{\rm iso}$) from naive counting arguments, which assume that
all contacts give rise to linearly independent constraints on
interparticle separations.  To understand this behavior, we decompose
the dynamical matrix $M=H-S$ for static packings of ellipsoidal
particles into two important components: the stiffness $H$ and stress
$S$ matrices.  We find that the stiffness matrix possesses $N(z_{\rm
iso} - z)$ eigenmodes ${\hat e}_0$ with zero eigenvalues even at
finite compression, where $N$ is the number of particles. In addition,
these modes ${\hat e}_0$ are nearly eigenvectors of the dynamical
matrix with eigenvalues that scale as $\Delta \phi$, and thus finite
compression stabilizes packings of ellipsoidal particles. At jamming
onset, the harmonic response of static packings of ellipsoidal particles
vanishes, and the total potential energy scales as $\delta^4$ for
perturbations by amplitude $\delta$ along these `quartic' modes,
${\hat e}_0$. These findings illustrate the significant differences
between static packings of spherical and ellipsoidal particles.
\end{abstract}


\maketitle

\section{Introduction}
\label{intro}

There have been many experimental~\cite{scott,bernal,gao2},
computational~\cite{berryman,torquato_prl,OHern2003}, and
theoretical~\cite{parisi,berthier} studies of the structural and
mechanical properties of disordered static packings of frictionless
disks in 2D and spheres in 3D.  In these systems, counting arguments,
which assume that all particle contacts give rise to linearly
independent impenetrability constraints on the particle positions,
predict that the minimum number of contacts required for the system to
be collectively jammed is $N_c \ge N_c^{\rm min} = N_{\rm dof} + 1$,
where $N_{\rm dof} = Nd$ for fixed boundary conditions and $N_{\rm
dof}=Nd - d$ for periodic boundary conditions~\cite{torquato,witten},
where $d$ is the spatial dimension and $N$ is the number of
particles~\cite{rattlers}. The additional contact is required because
contacts between hard particles provide only inequality constraints on
particle separations~\cite{torquato}.  In the large-system limit, this
relation for the minimum number of contacts reduces to $z \ge z_{\rm
iso}$, where $z=2N_c/N$ is the average contact number. Disordered
packings of frictionless spheres are typically isostatic at jamming
onset with $z = z_{\rm iso}$, and possess the minimal number of
contacts required to be collectively jammed~\cite{torquato}. Further,
it has been shown in numerical simulations that collectively jammed
hard-sphere packings correspond to mechanically stable soft-sphere
packings in the limit of vanishing particle
overlaps~\cite{OHern2003,makse,apphys}.

In contrast, several
numerical~\cite{torquato_ellipse,Mailman,zeravcic,delaney} and
experimental studies~\cite{chaikin_prl,chaikin_science} have found
that disordered packings of ellipsoidal particles possess fewer
contacts ($z < z_{\rm iso}$) than predicted by naive counting
arguments, which assume that all contacts give rise to linearly
independent constraints on the interparticle separations.  Despite
this, these packings were found to be mechanically stable
(MS)~\cite{torquato_ellipse,Mailman}.

In a recent manuscript~\cite{torquato_ellipse} by Donev, {\it et al.},
the authors explained this apparent contradiction---that static
packings of ellipsoidal particles are mechanically stable, yet possess
$z < z_{\rm iso}$.  The main points of the argument are included
here. The set of $N_c$ interparticle contacts imposes $N_c$
constraints, $f_{ij}=1-r_{ij}/\sigma_{ij} \le 0$, where $r_{ij}$ is
the center-to-center separation and $\sigma_{ij}$ is the contact
distance along ${\hat r}_{ij}$ between particles $i$ and $j$. In
disordered MS sphere packings with $z=z_{\rm iso}$, each of the $N_c$
interparticle contacts represents a linearly independent
constraint. In contrast, some of the $N_c$ contacts for MS packings of
ellipsoidal particles give rise to linearly {\it dependent}
constraints.  Linearly dependent constraints do not block the degrees
of freedom that appear in the constraint equations for sphere
packings, whereas they can block multiple degrees of freedom in
packings of ellipsoidal particles because they have convex particle
shape with a varying radius of curvature~\cite{torquato_ellipse}.

\begin{figure}[h]
\begin{center}
\includegraphics[width=0.45\textwidth]{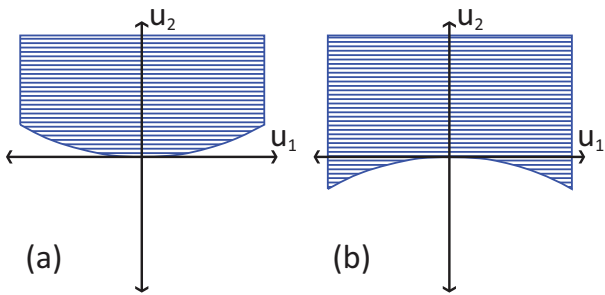}
\caption{Schematic diagram that illustrates locally (a) convex
(positive radius of curvature) and (b) concave (negative radius of
curvature) interparticle constraints in packings of hard ellipsoidal
particles. Inaccessible regions (with $f_{ij}>0$) are shaded blue. The
axes labeled $u_1$ and $u_2$ indicate two representative orthogonal
directions in configuration space.}
\label{convex_concave}
\end{center}
\end{figure}

For static packings of spherical particles, interparticle contacts
give rise to only ``convex" constraints (Fig.~\ref{convex_concave}
(a)), while contacts can yield ``convex" or ``concave" constraints in
packings of ellipsoidal particles (Figs.~\ref{convex_concave} (a) and
(b)). Note that the distinction between convex and concave constraints
is different than the distinction between convex and concave
particles. For example, ellipsoids have a convex particle shape, but
static ellipsoid packings can possess concave interparticle constraints.

%%%%%%%%%%%%%%%%%
%%%%%%%%%%%%%%%%%

In jammed packings of ellipsoidal particles, there are $N(z_{\rm iso}
- z)$ special directions $\hat{e}_0$ in configuration space along
which perturbations give rise to interparticle overlaps that scale
quadratically with the perturbation amplitude, $f_{ij} \sim
\delta^2$~\cite{torquato_ellipse,Mailman}. Displacements in all other
directions yield overlaps that scale linearly with $\delta$, $f_{ij}
\sim \delta$, as found for jammed sphere packings. This novel scaling
behavior for packings of ellipsoidal particles can be explained by
decomposing the dynamical matrix $M=H-S$ for these packings into two
important components: the stiffness matrix $H$ that contains all
second-order derivatives of the total potential energy $V$ with
respect to the configurational degrees of freedom, and the stress
matrix $S$ that includes all first-order derivatives of $V$ with
respect to the particle coordinates.  The directions $\hat{e}_0$ are
the eigenvectors of the stiffness matrix $H$ with zero eigenvalues.

For static packings of ellipsoidal particles at the jamming threshold
($\Delta\phi=0$) that interact via purely repulsive linear spring
potentials ({\it i.e.} $V\sim f_{ij}^2$), we find that the total
potential energy increases quartically when the system is perturbed by
$\delta$ along the $\hat{e}_0$ directions, $V\propto c\delta^4$, where
the constant $c>0$. Also, at the jamming threshold, the stress matrix
$S=0$ and zero modes of the stiffness matrix are zero modes of the
dynamical matrix.

In this manuscript, we will investigate how the mechanical stability
of packings of ellipsoidal particles is modified at finite
compression ($\Delta \phi > 0$).  For example, when a system at finite
$\Delta \phi$ is perturbed by amplitude $\delta$ along $\hat{e}_0$, do
quadratic terms in $\delta$ arise in the total potential energy or do the
contributions remain zero to second order?  If quadratic terms are
present, do they stabilize or destabilize the packings, and how do the 
lowest frequency modes of the dynamical matrix scale with $\Delta \phi$ and
aspect ratio?

We find a number of key results for static packings of ellipsoidal
particles at finite compression ($\Delta \phi > 0$) including: 1)
Packings of ellipsoidal particles generically satisfy $z < z_{\rm
iso}$~\cite{Mailman,zeravcic,torquato_ellipse,delaney}; 2) The
stiffness matrix $H$ possesses $N(z_{\rm iso} - z)$ eigenmodes ${\hat
e}_0$ with zero eigenvalues even at finite compression $(\Delta \phi >
0)$; and 3) The modes ${\hat e}_0$ are nearly eigenvectors of the
dynamical matrix (and the stress matrix $-S$) with eigenvalues that
scale as $c \Delta \phi$, with $c>0$, and thus finite compression
stabilizes packings of ellipsoidal particles~\cite{Mailman}.  In
contrast, for static packings of spherical particles, the stiffness
matrix contributions to the dynamical matrix stabilize all 
modes near jamming onset. At jamming onset ($\Delta \phi =0$),
the harmonic response of packings of ellipsoidal particles vanishes,
and the total potential energy scales as $\delta^4$ for perturbations
by amplitude $\delta$ along these `quartic' modes, ${\hat e}_0$. Our
findings illustrate the significant differences between amorphous
packings of spherical and ellipsoidal particles.

The remainder of the manuscript will be organized as follows. In
Sec.~\ref{methods}, we describe the numerical methods that we employed
to measure interparticle overlaps, generate static packings, and
assess the mechanical stability of packings of ellipsoidal
particles. In Sec.~\ref{vibration}, we describe results from
measurements of the density of vibrational modes in the harmonic
approximation, the decomposition of the dynamical matrix eigenvalues
into contributions from the stiffness and stress matrices, and the
relative contributions of the translational and rotational degrees of
freedom to the vibrational modes as a function of overcompression and
aspect ratio using several packing-generation protocols. In
Sec.~\ref{conclusion}, we summarize our conclusions and provide
promising directions for future research. We also include two
appendices. In Appendix~\ref{contact_formation}, we show that the
formation of new interparticle contacts affects the scaling behavior
of the potential energy with the amplitude of small perturbations
along eigenmodes of the dynamical matrix.  In
Appendix~\ref{appendixb}, we provide analytical expressions for the
elements of the dynamical matrix for ellipse-shaped particles in 2D
that interact via a purely repulsive linear spring potential.

\begin{figure}
\begin{center}
\includegraphics[width=0.3\textwidth]{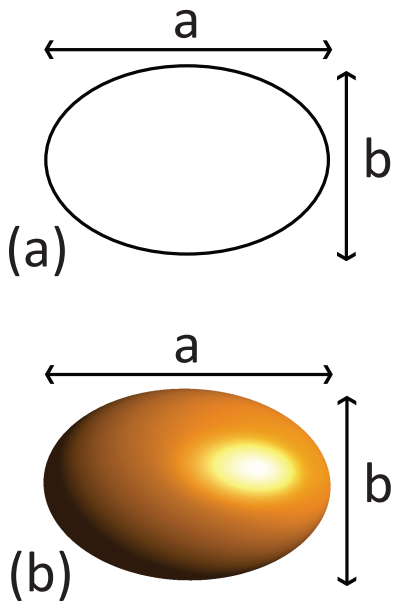}
\caption{We focus on (a) ellipses in 2D with aspect ratio $\alpha =
a/b$ defined as the ratio of the major to minor axis and (b) prolate
ellipsoids in 3D where $\alpha$ is the ratio of the polar to
equatorial lengths.}
\label{intro_fig1}
\end{center}
\end{figure} 

\section{Methods}
\label{methods}

In this section, we describe the computational methods employed to
generate static packings of convex, anisotropic particles, {\it i.e.}
ellipses in 2D and prolate ellipsoids in 3D with aspect ratio $\alpha
= a/b$ of the major to minor axes (Fig.~\ref{intro_fig1}), and analyze
their mechanical properties.  To inhibit ordering in 2D, we studied
bidisperse mixtures (2-to-1 relative number density), where the ratio
of the major (and minor) axes of the large and small particles is
$a_l/a_s=b_l/b_s=1.4$. In 3D, we focused on a monodisperse size
distribution of prolate ellipsoids.  We employed periodic boundaries
conditions in unit square (2D) and cubic (3D) cells and studied
systems sizes in the range from $N=30$ to $960$ particles to address
finite-size effects.

%%%%%%%%%%%%%%%%%
\begin{figure}
\begin{center}
\includegraphics[width=0.3\textwidth]{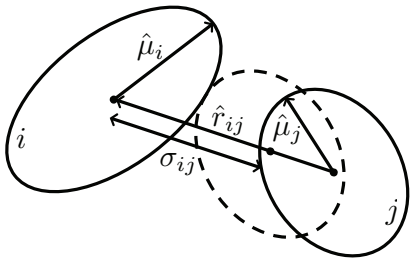}
\caption{Definition of the 
contact distance $\sigma_{ij}$ for ellipsoidal particles $i$ and $j$
with unit vectors ${\hat \mu}_i$ and ${\hat \mu}_j$ that characterize
the orientations of their major axes.  $\sigma_{ij}$ is the
center-to-center separation at which ellipsoidal particles first
touch when they are brought together along ${\hat r}_{ij}$
at fixed orientation.}
\label{fig:Fig10_chapter} 
\end{center}
\end{figure}
%%%%%%%%%%%%%%%%%

\subsection{Contact distance}
\label{contact}

In both 2D and 3D, we assume that particles interact via the following
pairwise, purely repulsive linear spring potential
\begin{equation}
V(r_{ij}) = \begin{cases} \frac{\epsilon}{2}\left(1-\frac{r_{ij}}{\sigma_{ij}}\right)^2 & r_{ij} \le \sigma_{ij} \\
0 & r_{ij} > \sigma_{ij}, 
\end{cases}
\label{eq:ellipse_energy}
\end{equation}
where $\epsilon$ is the characteristic energy of the interaction,
$r_{ij}$ is the center-to-center separation between particles $i$ and
$j$, and $\sigma_{ij}$ is the orientation-dependent center-to-center
separation at which particles $i$ and $j$ come into contact as shown
in Fig.~\ref{fig:Fig10_chapter}.  Below, energies, lengths, and time
scales will be expressed in units of $\epsilon$, $l=\sqrt{I/m}$, and
$l\sqrt{m/\epsilon}$, respectively, where $m$ and $I$ are the mass and
moment of inertia of the ellipsoidal particles.

Perram and Wertheim developed an efficient method for calculating the
exact contact distance between ellipsoidal particles with any aspect ratio and
size distribution in 2D and 3D~\cite{Perram}. In their formulation,
the contact distance is obtained from 
\begin{eqnarray}
\label{sigma_functional}
\sigma_{ij}&=&\min_{\lambda} \sigma_{ij}(\lambda),\\\nonumber
\sigma_{ij}(\lambda)&=&
\frac{\sigma^0_{ij}(\lambda)}
{\sqrt{1-\frac{\chi(\lambda)}{2}\displaystyle\sum_{\pm}
\frac{(\beta(\lambda)\hat{r}_{ij}\cdot\hat{\mu}_i\pm
\beta(\lambda)^{-1}\hat{r}_{ij}\cdot\hat{\mu}_j)^2}
{1\pm\chi(\lambda)\hat{\mu}_i\cdot\hat{\mu}_j}}},\\\nonumber
\sigma_{ij}^0(\lambda)&=&\frac{1}{2} \sqrt{
\frac{b_i^{2}}{\lambda}+ \frac{b_j^{2}}{1- \lambda} },\\\nonumber
\chi(\lambda)
&=&\left(\frac{\left(a_{i}^{2}-b_{i}^{2}\right) \left(a_{j}^{2}-
b_{j}^{2}\right)} {\left(a_{j}^{2}+\frac{1-\lambda}{\lambda}
b_{i}^{2}\right)
\left(a_{i}^{2}+\frac{\lambda}{1-\lambda}b_{j}^{2}\right)}
\right)^{1/2},\\\nonumber
\beta(\lambda)&=&
\left(\frac{\left(a_i^2-b_i^2\right)\left(a_j^2+\frac{1-\lambda}
{\lambda}b_i^2\right)}
{\left(a_j^2-b_j^2\right)\left(a_i^2+\frac{\lambda}{1-
\lambda}b_j^2\right)}\right)^{1/4}.
\end{eqnarray} 

\begin{figure}
\begin{center}
\includegraphics[width=0.45\textwidth]{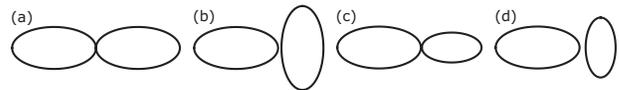}
\caption{Ellipses with $\alpha=2$ positioned at the Gay-Berne
contact distance $\sigma^a_{ij}$. For two ellipses with the
same size, the (a) end-to-end configuration is exact, while the (b)
side-to-end configuration has a $5\%$ relative error. For two
ellipses with $a_j/a_i=1.4$, the (c) end-to-end configuration has a relative
error of $1\%$, while the (d) side-to-end configuration has a relative error
of $10\%$.}
\label{fig:Fig12_chapter} 
\end{center}
\end{figure}

The approximation $\sigma^a_{ij}=\sigma_{ij}(\lambda=1/2)$ is
equivalent to the commonly used Gay-Berne approximation for the contact
distance~\cite{Perram1996,Cleaver}.  The accuracy of the Gay-Berne
approximation depends on the relative orientation of the two
ellipsoidal particles, and in general is more accurate for
monodisperse systems. For example, in Fig.~\ref{fig:Fig12_chapter}, we
show $\sigma_{ij}^a$ for several relative orientations of both
monodisperse and bidisperse systems. The relative deviation from the
true contact distance can be as large as $e\sim 10\%$ for
$a_j/a_i=1.4$ and $\alpha=2$.  Thus, the Gay-Berne
approximation should be used with caution when studying polydisperse
packings of ellipsoidal particles.  For monodisperse
ellipses with $\alpha=2$, $0\% < e < 5\%$. We find similar results
for 3D systems. Unless stated otherwise, we employ the exact
expression for contact distance, and thus $\sigma_{ij} =\sigma_{ij}
(\lambda_{\rm min})$, $\beta = \beta(\lambda_{\rm min})$, $\chi =
\chi(\lambda_{\rm min})$, and $\sigma_{ij}^0
=\sigma_{ij}^0(\lambda_{\rm min})$, where $\lambda_{\rm min}$ is the
minimum obtained from Eq.~\ref{sigma_functional}.

%%%%%%%%%%%%%%%%%
%%%%%%%%%%%%%%%%%

\subsection{Packing generation protocol}
\label{packing}

We employ a frequently used isotropic compression method for soft,
purely repulsive particles~\cite{xu,gao} to generate static packings
of ellipsoidal particles at jamming onset ($\Delta \phi=0$).  Static
packings at jamming onset are characterized by infinitesimal but
nonzero total potential energy and pressure.  This isotropic
compression method consists of the following steps.  We begin by
randomly placing particles at low packing fraction ($\phi_0=0.2$) with
random orientations and zero velocities in the simulation cell.  We
successively compress the system by small packing fraction increments
$\delta \phi = 10^{-3}$, with each compression followed by conjugate
gradient (CG) energy minimization until the total potential energy per
particle drops below a small threshold, $V/N \le V_{\rm tol} =
10^{-16}$, or the total potential energy per particle between
successive iterations of the minimization routine is $(V_{t+1} -
V_{t})/V_t \le V_{\rm tol}$.  The algorithm switches from compression
to decompression if the minimized energy is greater than $2 V_{\rm
tol}$.  Each time the algorithm toggles from compression to
decompression or vice versa, the packing fraction increment is halved.

The packing-generation algorithm is terminated when the total
potential energy per particle satisfies $V_{\rm tol} < V/N < 2 V_{\rm
tol}$.  Thus, using this method we can locate the jammed packing
fraction $\phi_J$ and particle positions at jamming onset for each initial
condition to within $10^{-8}$.  After jamming onset is identified, we
also generate configurations at specified $\Delta \phi = \phi-\phi_J$
over six orders of magnitude from $10^{-8}$ to $10^{-2}$ by applying a
prescribed set of compressions with each followed by energy
minimization.

To determine whether the accuracy of the energy minimization algorithm
affects our results (see Sec.~\ref{protocol}), we calculate the
eigenvalues of the dynamical matrix as a function of the total kinetic
energy (or deviation from zero in force and torque balance on each
particle) at each $\Delta \phi$.  To do this, we initialize the system
with MS packings from the above packing-generation algorithm and use
molecular dynamics (MD) simulations with damping terms proportional to
the translational and rotational velocities of the ellipsoidal
particles to remove excess kinetic energy from the
system~\cite{foot_md}.  The damped MD simulations are terminated when
the total kinetic energy per particle is below $K/N = K_{\rm tol}$,
where $K_{\rm tol}$ is varied from $10^{-16}$ to $10^{-32}$. This
provides accuracy in the particle positions of the energy minimized
states in the range from $10^{-8}$ to $10^{-16}$.

\begin{figure}
\begin{center}
\includegraphics[width=0.3\textwidth]{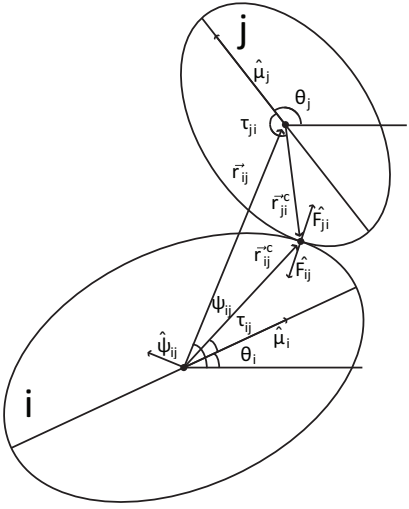}
\caption{Geometry of two ellipses in contact. The angles $\theta_i$
and $\theta_j$ characterize the orientation of particles $i$ and $j$
relative to the horizontal axis, {\it i.e.} ${\hat \mu}_i = \cos
\theta_i {\hat x} + \sin \theta_i {\hat y}$.  $\psi_{ij}$ gives the
angle between the center-to-center separation vector ${\vec r}_{ij}$
and the horizontal axis and ${\hat \psi}_{ij} = -\sin \psi_{ij} {\hat
x} +\cos \psi_{ij} {\hat y}$ is the angle unit vector in polar
coordinates. The unit vector ${\hat F}_{ij} = - {\hat F}_{ji}$ points
in the direction of the force on particle $i$ due to particle $j$ at
the point of contact. ${\vec r}^c_{ij}$ points from the center of
particle $i$ to the point of contact with particle $j$, and
$\tau_{ij}$ is the angle between ${\hat \mu}_i$ and ${\vec
r}^c_{ij}$.}
\label{torque_schematic}
\end{center}
\end{figure}

For the damped MD simulations, we solve Newton's equations of motion
(using fifth-order Gear predictor-corrector methods~\cite{allen}) for
the center of mass position and angles that characterize the
orientation of the long axis of the ellipsoidal particles.  In 2D, we
solve
\begin{eqnarray}
\label{newton}
m\frac{d^2{\vec r}_i}{dt^2} & = & \sum_{i>j} {\vec F}_{ij} -b_r {\vec
  v}_i\\ I \frac{d^2 \theta_i}{dt^2} & = & \sum_{i>j} T_{ij} -
b_{\theta} {\dot \theta}_i,
\label{angle}
\end{eqnarray} 
where $\theta_i$ is the angle the long axis of ellipse $i$ makes with
the horizontal axis, ${\vec v}_i$ is the translational velocity of
particle $i$, ${\dot \theta}_i$ is the rotational speed of particle
$i$, $b_r$ and $b_{\theta}$ are the damping coefficients for the
position and angle degrees of freedom, and the moment of inertia
$I=m(a^2+b^2)/4$. The force ${\vec F}_{ij}$ on ellipse $i$ arising
from an overlap with ellipse $j$ is
\begin{eqnarray}
\label{fijvec}
\vec{F}_{ij}&=&-F_{ij}\frac{\hat{r}_{ij}-\frac{\partial \ln \sigma_{ij} }{\partial\psi_{ij}}\hat{\psi}_{ij}}
{\sqrt{1+\big(\frac{\partial \ln \sigma_{ij} }{\partial\psi_{ij}}\big)^2}},
\label{eq:force}
\end{eqnarray}
where 
\begin{eqnarray}
\label{fij}
F_{ij}&=&\sqrt{1+\left(\frac{\partial \ln \sigma_{ij} }{\partial\psi_{ij}}\right)^2}
\left|\frac{\partial V(r_{ij})}{\partial r_{ij}}\right|,\\
\frac{\partial \ln \sigma_{ij}}{\partial \psi_{ij}}&=&-\frac{\chi}{2}\frac{\sigma_{ij}^2}{(\sigma_{ij}^0)^2}
\big[(\beta^2-\chi)\sin\left[2(\psi_{ij}-\theta_i)\right]+\nonumber\\
&&(\beta^{-2}-\chi)\sin\left[2(\psi_{ij}-\theta_j)\right]\big] \times \nonumber\\
&&\left(1-\chi^2\cos^2\left[\theta_i-\theta_j\right]\right)^{-1},
\label{eq:force2}
\end{eqnarray}
$-dV(r_{ij})/dr_{ij}=\sigma^{-1}_{ij}(1-\frac{r_{ij}}{\sigma_{ij}})$
for the purely repulsive linear spring potential in
Eq.~\ref{eq:ellipse_energy}, and $\hat{r}_{ij}$ and
$\hat{\psi}_{ij}$ are illustrated in Fig.~\ref{torque_schematic}.

To calculate the torque $T_{ij}=[\vec{r_{ij}^c}\times\vec{F}_{ij}]
\cdot {\hat z}$ in Eq.~\ref{angle}, we must identify the point
of contact between particles $i$ and $j$,
\begin{eqnarray}
\label{moment}
\vec{r}_{ij}^c&=&\frac{b_i}{2}\frac{1}{\sqrt{\alpha^{-2}+\tan^2 \tau_{ij} }}\times\nonumber\\
&&\big[(\cos \theta_i - \sin \theta_i \tan \tau_{ij} )\hat{x}+\nonumber\\
&&(\sin \theta_i \tan \tau_{ij} + \cos \theta_i )\hat{y}\big],
\end{eqnarray}
where
\begin{eqnarray}
\tan \tau_{ij} &=&\alpha^{-2} \frac{\tan(\psi_{ij}-\theta_i)-
\frac{\partial\ln\sigma_{ij}}{\partial\psi_{ij}}}{1+\tan(\psi_{ij}-\theta_i)
\frac{\partial\ln\sigma_{ij}}{\partial\psi_{ij}}}
\label{eq:point_of_contact}
\end{eqnarray}
and $\vec{r}_{ij}^c$, $\psi_{ij}$, and $\tau_{ij}$ are depicted in
Fig.~\ref{torque_schematic}. From Eqs.~\ref{eq:force} and~\ref{moment},
we find
\begin{eqnarray}
T_{ij}&=&-\frac{b_iF_{ij}}{2} \times \nonumber \\
& & \frac{\big(1-\alpha^{-2}\big)\tan
\tau_{ij}} {\sqrt{\big(1+\alpha^{-4}\tan^2 \tau_{ij}
\big)\big(1+\alpha^{-2}\tan^2 \tau_{ij}\big)}}.
\label{eq:torque}
\end{eqnarray}

\subsection{Dynamical matrix calculation}
\label{DM_method}

To investigate the mechanical properties of static packings of
ellipsoidal particles, we will calculate the eigenvalues of the
dynamical matrix and the resulting density of vibrational modes in the
harmonic approximation~\cite{schreck2011}.  The dynamical matrix is defined as
\begin{eqnarray}
M_{kl}&=&\frac{\partial^2V}{\partial u_k\partial u_l},
\label{eq:DM_eq}
\end{eqnarray}
where $u_k$ (with $k=1,\ldots,d_f N$) represent the $d_f N$ degrees of
freedom in the system and $d_f$ is the number of degrees of freedom
per particle. In 2D $d_f = 3$ with ${\vec u} =
\{x_1$,$x_2$,$\ldots$,$x_N$,$y_1$,$y_2$,$\ldots$,$y_N$,$l_2
\theta_1$,$l_2 \theta_2$,$\ldots$,$l_2\theta_N$\} and in 3D for
prolate ellipsoids $d_f = 5$ with ${\vec u} =
\{x_1$,$x_2$,$\ldots$,$x_N$,$y_1$,
$y_2$,$\ldots$,$y_N$,$z_1$,$z_2$,$\ldots$,$z_N$,$l_{\theta}^1
\theta_1$,$l_{\theta}^2
\theta_2$,$\ldots$,$l_{\theta}^N\theta_N$,$l_{3} \phi_1$,$l_{3}
\phi_2$,$\ldots$, $l_{3}\phi_N$\}, where $\theta_i$ is the polar angle
and $\phi_i$ is the azimuthal angle in spherical coordinates,
$l_2=\sqrt{a^2+b^2}/2$, $l_3=\sqrt{\big(a^2+b^2\big)/5}$, and
$l_{\theta}^i=\sqrt{\big(2b^2+(a^2-b^2)\sin^2 \phi_i\big)/5}$.

The dynamical matrix requires calculations of the first and second
derivatives of the total potential energy $V$ with respect to all
positional and angular degrees of freedom in the system.  The first
derivatives of $V$ with respect to the positions of the centers of
mass of the particles ${\vec r}_i$ can be obtained from
Eq.~\ref{fijvec}. In 2D, there is only one first derivative involving
angles, $F_{\theta}^i=-\partial V(r_{ij})/\partial\theta_i$, where
\begin{eqnarray}
\label{angular_deriv}
F_{\theta}^i&=&\frac{1}{4}\chi\bigg(\frac{\sigma_{ij}}{\sigma_{ij}^0}\bigg)^2(2\alpha A(B_+ + B_-)+\chi C(B_+^2-B_-^2)),\\
A&=&\frac{y_{ij}\cos \theta_i -x_{ij}\sin \theta_i }{r_{ij}}, \nonumber \\
B_{\pm}&=&\frac{\alpha(x_{ij} \cos \theta_i + y_{ij}\sin \theta_i )\pm \alpha^{-1}(x_{ij}\cos \theta_j +y_{ij}\sin \theta_j) }
{(1+\chi\cos[\theta_i-\theta_j])r_{ij}},\nonumber \\
C&=&\cos^2(\theta_i-\theta_j). \nonumber
\end{eqnarray}
Complete expressions for the matrix elements of the dynamical matrix
for ellipses in 2D are provided in Appendix~\ref{appendixb}.  In 3D,
we calculated the first derivatives of $V$ with respect to the
particle coordinates analytically, and then evaluated the second
derivatives for the dynamical matrix numerically.

The vibrational frequencies in the harmonic approximation can be
obtained from the $Nd_f - d$ nontrivial eigenvalues $m_i$ of the
dynamical matrix, $\omega_i = \sqrt{m_i/\epsilon} b_s$.  $d$ of the
eigenvalues are zero due to periodic boundary conditions.  For all
static packings, we have verified that the smallest nontrivial
eigenvalue satisfies $m_{\rm min}/N>10^{-10}$.

Below, we will study the density of vibrational frequencies $D(\omega)
= (N(\omega+\Delta\omega)-N(\omega))/(N_{\rm dof} \Delta\omega)$ as a
function of compression $\Delta \phi$ and aspect ratio $\alpha$, where
$N(\omega)$ is the number of vibrational frequencies less than
$\omega$. We will also investigate the relative contributions of the
translational and rotational degrees of freedom to the nontrivial
eigenvectors of the dynamical matrix, ${\hat m}_i =
\{m_{xi}^{j=1},m_{yi}^{j=1},m_{\theta i}^{j=1},\ldots,m_{xi}^{j=N},
m_{yi}^{j=N},m_{\theta i}^{j=N}\}$ for ellipses in 2D and ${\hat m}_i
= \{m_{xi}^{j=1},m_{yi}^{j=1},m_{zi}^{j=1},m_{\theta i}^{j=1},m_{\phi
i}^{j=1},\ldots,m_{xi}^{j=N},m_{yi}^{j=N},m_{z i}^{j=N},$ $m_{\theta
i}^{j=N},m_{\phi i}^{j=N}\}$ for prolate ellipsoids in 3D, where $i$
labels the eigenvector and runs from $1$ to $Nd_f - d$. The
eigenvectors are normalized such that ${\hat m}_i^2 = 1$.

\subsection{Dynamical matrix decomposition}
\label{DM_decomp}

The dynamical matrix (Eq.~\ref{eq:DM_eq}) can be decomposed
into two component matrices $M=H-S$: 1) the stiffness matrix $H$ that includes
only second-order derivatives of the total potential energy $V$ with
respect to the configurational degrees of freedom and 2) the stress
matrix $S$ that includes only first-order derivatives of $V$.  The
$kl$ elements of $H$ and $S$ are given by
\begin{eqnarray}
H_{kl}&=&\sum_{i>j}
\frac{\partial^2V}{\partial(r_{ij}/\sigma_{ij})^2}
\frac{\partial(r_{ij}/\sigma_{ij})}{\partial u_k}
\frac{\partial(r_{ij}/\sigma_{ij})}{\partial u_l}\\
S_{kl}&=&-\sum_{i>j}\frac{\partial V}{\partial(r_{ij}/\sigma_{ij})}
\frac{\partial^2(r_{ij}/\sigma_{ij})}{\partial u_k\partial u_l},
\end{eqnarray}
where the sums are over distinct pairs of overlapping particles $i$
and $j$. Since $\partial^2V/\partial(r_{ij}/\sigma_{ij})^2=\epsilon$
for the purely repulsive linear spring potential
(Eq.~\ref{eq:ellipse_energy}), the stiffness matrix depends only on
the geometry of the packing ({\it i.e.}
$\partial(r_{ij}/\sigma_{ij})/\partial u_k)$. Also, at zero
compression $\Delta\phi=0$, $S=0$, $M=H$, and only the stiffness
matrix contributes to the dynamical matrix.  The frequencies
associated with the eigenvalues $h_i$ of the stiffness matrix (at any
$\Delta \phi$) are denoted by $\omega_{hi} = \sqrt{h_i/\epsilon} b_s$, 
and the stiffness matrix eigenvectors are normalized such that 
${\hat h}_i^2=1$.

\subsection{Contact number}
\label{contact_number}

When counting the number of interparticle contacts $N_c$, we remove
all rattler particles (defined as those with fewer than $d+1$
contacts) and do not include the contacts that rattler particles make
with non-rattler particles~\cite{foot_contact}. Removing these
contacts may cause non-rattler particles to become rattlers, and thus
this process is performed recursively.  Note that for ellipsoidal
particles with $d+1$ contacts, the lines normal to the points (or
planes in 3D) of contact must all intersect, otherwise the system is
not mechanically stable.  The number of contacts per particle is
defined as $z=N_c/(N-N_r)$, where $N_r$ is the number of rattlers.  We
find that the number of rattler particles decreases with aspect ratio
from approximately $5\%$ of the system at $\alpha = 1$ to zero for
$\alpha > 1.2$ in both 2D and 3D.
   
\section{Results}
\label{vibration}

\begin{figure}[h]
\begin{center}
\includegraphics[width=0.45\textwidth]{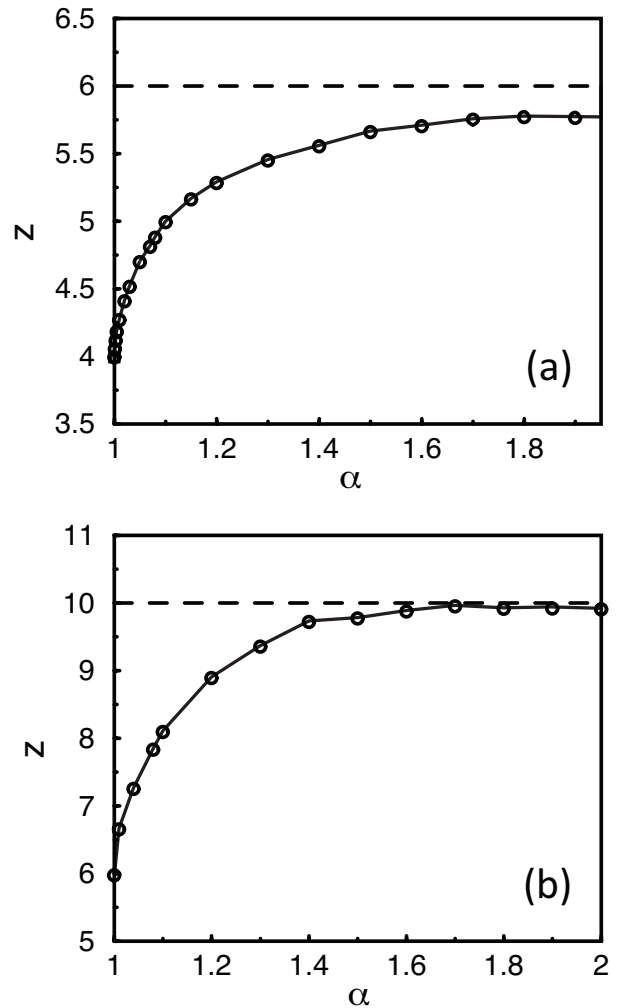}
\caption{Average contact number $z$ versus aspect ratio $\alpha$ for
static packings of (a) bidisperse ellipses in 2D and (b)
prolate ellipsoids in 3D at jamming onset. The isostatic
values $z_{\rm iso} = 6$ (2D) and $10$ (3D) are indicated by dashed
lines.}
\label{z}
\end{center}
\end{figure}

Static packings of ellipsoidal particles at jamming onset typically
possess fewer contacts than predicted by isostatic counting
arguments~\cite{torquato_ellipse}, $z < z_{\rm iso}$, over a wide
range of aspect ratio as shown in Fig.~\ref{z}.  This finding raises a
number of important questions. For example, are static packings of
ellipsoidal particles mechanically stable at finite $\Delta \phi > 0$,
{\it i.e.} does the dynamical matrix for these systems possess nontrivial
zero-frequency modes at $\Delta \phi > 0$?  In this section, we will
show that packings of ellipsoidal particles are indeed mechanically
stable (with no nontrivial zero-frequency modes) by calculating the
dynamical, stress, and stiffness matrices for these systems as a
function of compression $\Delta \phi$, aspect ratio $\alpha$, and
packing-generation protocol.  Further, we will show that the density
of vibrational modes for these systems possesses three characteristic
frequency regimes and determine the scaling of these characteristic
frequencies with $\Delta \phi$ and $\alpha$.

%%%%%%%%%%%%%%%%%
%%%%%%%%%%%%%%%%%

\subsection{Density of vibrational frequencies $D(\omega)$}
\label{DOS}

A number of studies have shown that amorphous sphere packings are
fragile solids in the sense that the density of vibrational
frequencies (in the harmonic approximation) $D(\omega)$ for these
systems possesses an excess of low-frequency modes over Debye solids
near jamming onset, {\it i.e.} a plateau forms and extends to lower
frequencies as $\Delta \phi \rightarrow
0$~\cite{OHern2003,silbert2005,wyart}.  In this work, we will
calculate $D(\omega)$ as a function of $\Delta \phi$ and aspect ratio
$\alpha$ for amorphous packings of ellipsoidal particles and show that
the density of vibrational modes for these systems shows significant
qualitative differences from that for spherical particles.

\begin{figure}
\includegraphics[width=0.4\textwidth]{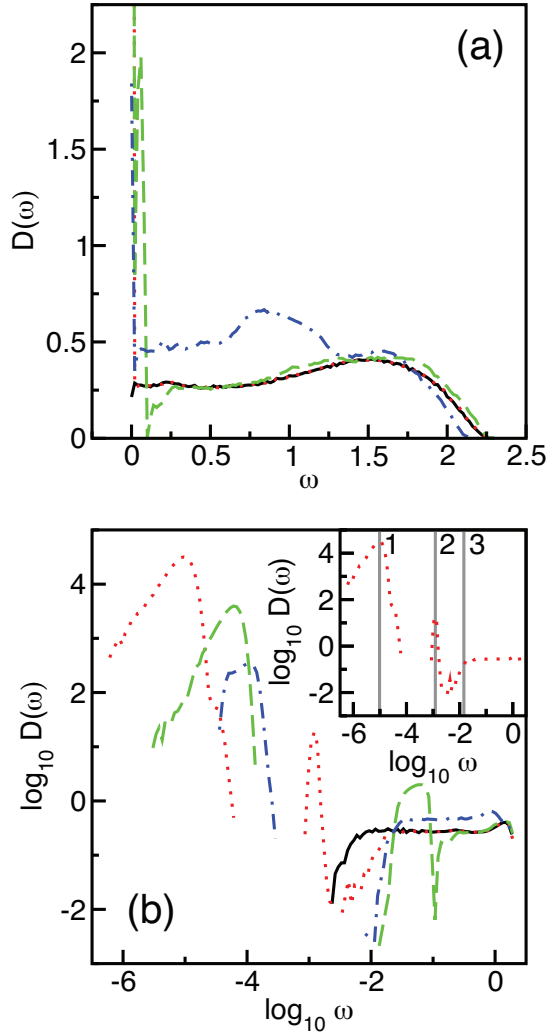}
\caption{(a) The density of vibrational frequencies $D(\omega)$ for
$N=240$ ellipse-shaped particles at $\Delta\phi=10^{-8}$ with aspect
ratio $\alpha=1.0$ (solid), $1.001$ (dotted), $1.05$ (dashed), and
$2.0$ (dot-dashed). $D(\omega)$ for $\alpha=1$ has been scaled by
$2/3$ relative to those with $\alpha>1$ to achieve collapse at low
aspect ratios. (b) $D(\omega)$ for the same aspect ratios in (a) on a
$\log$-$\log$ scale.  The inset illustrates the three characteristic
frequencies $\omega_1$, $\omega_2$, and $\omega_3$ in $D(\omega)$ for
$\alpha=1.001$.}
\label{fig_DM_2D}
\end{figure}

\begin{figure}
\includegraphics[width=0.45\textwidth]{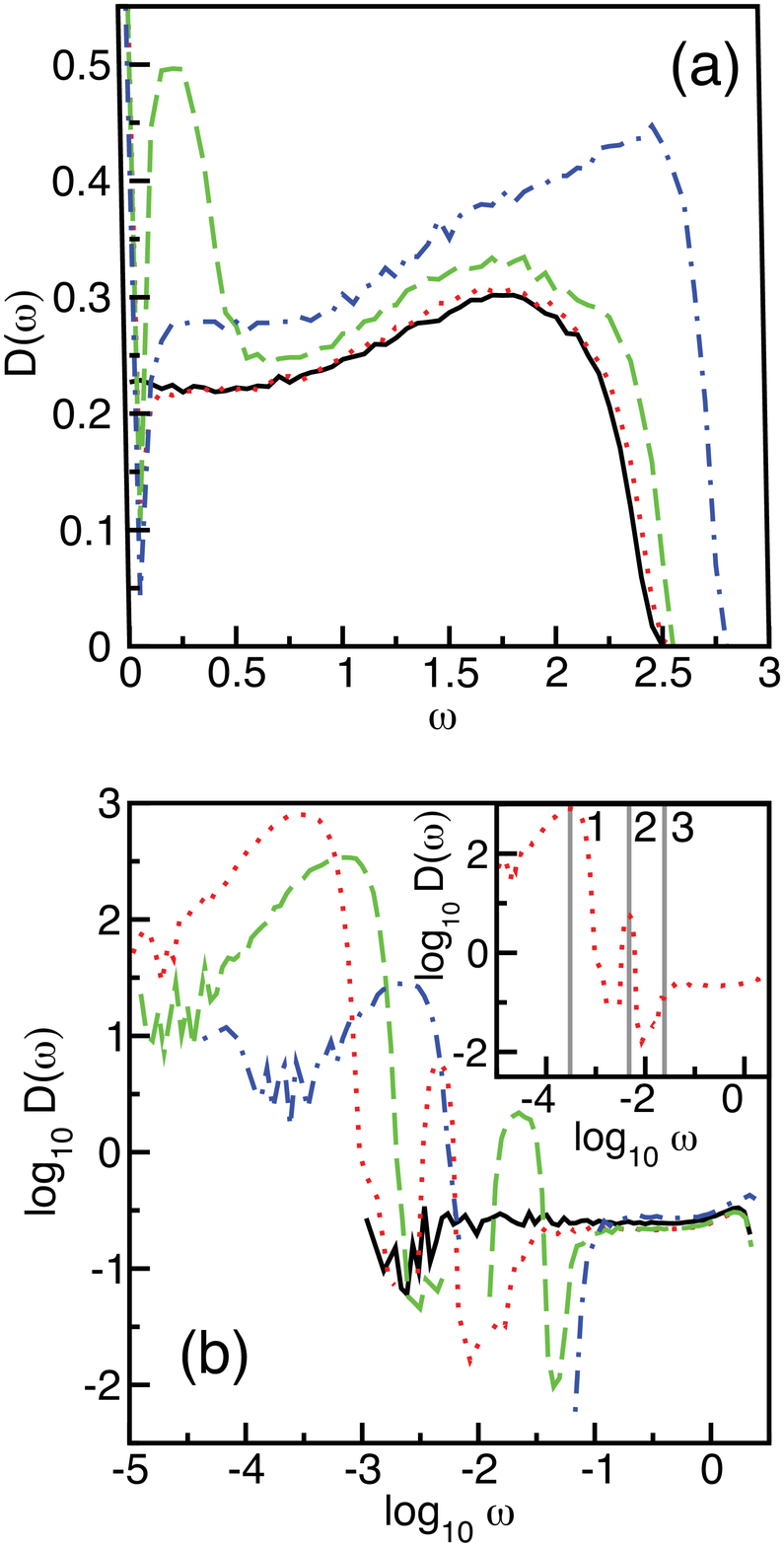}
\caption{(a) The density of vibrational frequencies $D(\omega)$ for
$N=512$ prolate ellipsoids at $\Delta\phi=10^{-6}$ for $\alpha=1.0$
(solid), $1.001$ (dotted), $1.005$ (dashed), and $1.2$ (dot-dashed).
$D(\omega)$ for $\alpha=1$ has been scaled by $3/5$ relative to those
with $\alpha>1$ to achieve collapse at low aspect ratios. (b)
$D(\omega)$ for the same aspect ratios in (a) on a $\log$-$\log$
scale. The inset illustrates the three characteristic frequencies $\omega_1$, 
$\omega_2$, and $\omega_3$ in $D(\omega)$ for $\alpha=1.001$.}
\label{fig_DM_3D}
\end{figure}

In Figs.~\ref{fig_DM_2D} (a) and (b), we show $D(\omega)$ on linear
and $\log$-$\log$ scales, respectively, for ellipse-shaped particles
in 2D at $\Delta \phi = 10^{-8}$ over a range of aspect ratios from
$\alpha = 1$ to $2$.  We find several key features in $D(\omega)$: 1)
For low aspect ratios $\alpha < 1.05$, $D(\omega)$ collapses with that
for disks ($\alpha=1$) at intermediate and large frequencies $0.25 <
\omega < 2.25$; 2) For large aspect ratios $\alpha \ge 2$, $D(\omega)$
is qualitatively different for ellipses than for disks over the entire
frequency range; and 3) A strong peak near $\omega = 0$ and a smaller
secondary peak at intermediate frequencies (evident on the log-log
scale in Fig.~\ref{fig_DM_2D} (b)) occur in $D(\omega)$ for $\alpha >
1$. Note that at finite compression $\Delta \phi > 0$, we do not find
any nontrivial zero-frequency modes of the dynamical matrix in static
packings of ellipses and ellipsoids. The only zero-frequency modes in
these systems correspond to the $d$ constant translations that arise
from periodic boundary conditions and zero-frequency modes associated
with `rattler' particles with fewer than $d+1$ interparticle contacts.

\begin{figure}
\includegraphics[width=0.4\textwidth]{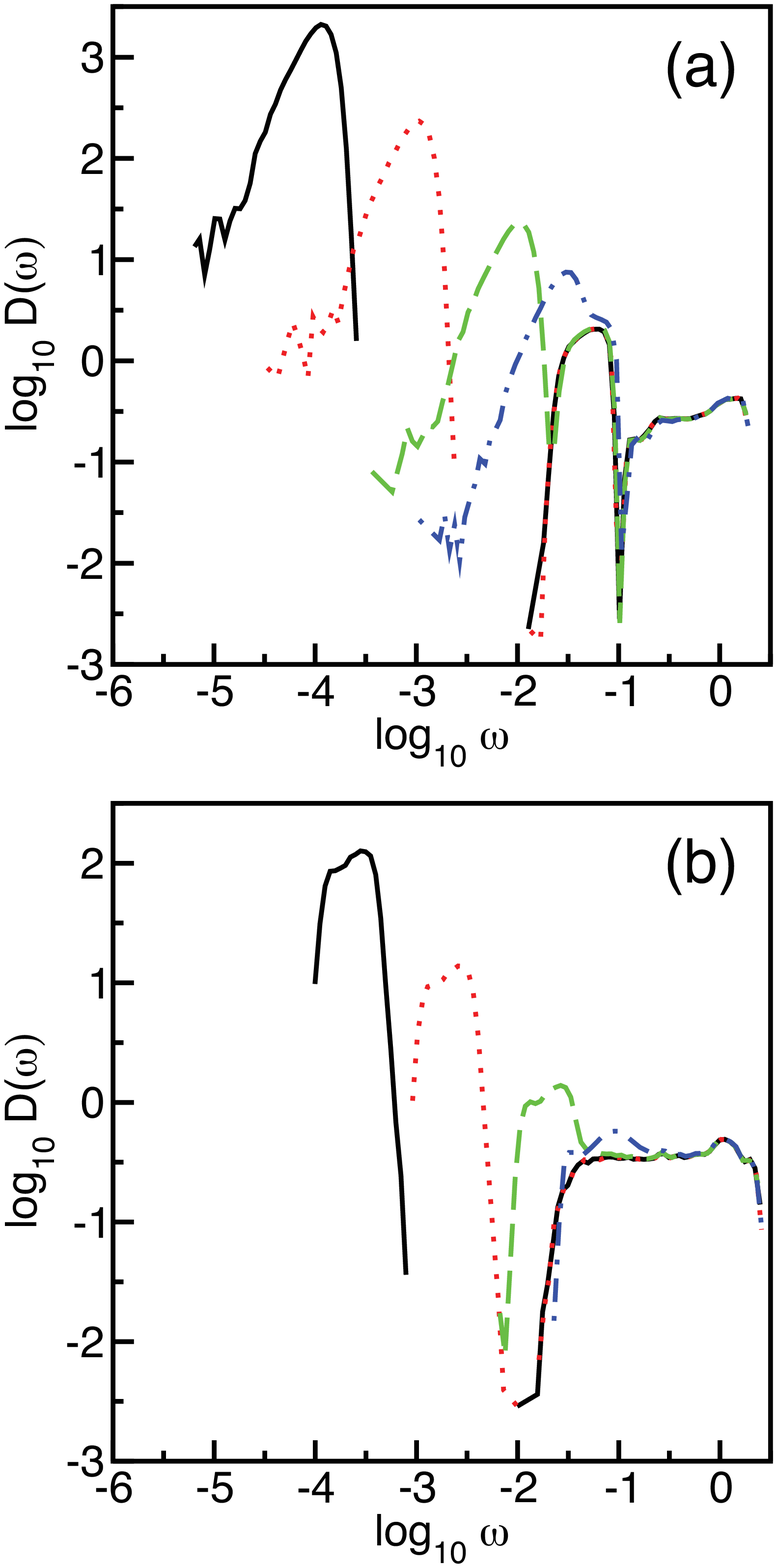}
\caption{The density of vibrational frequencies $D(\omega)$ for $N=240$ 
ellipses as a function 
of compression $\Delta\phi=10^{-7}$ (solid), $10^{-5}$ (dashed),
$10^{-3}$ (dotted), and $10^{-2}$ (dot-dashed) for
(a) $\alpha=1.05$ and (b) $2$.}
\label{fig_DM_2D_comp}
\end{figure}

\begin{figure}
\includegraphics[width=0.4\textwidth]{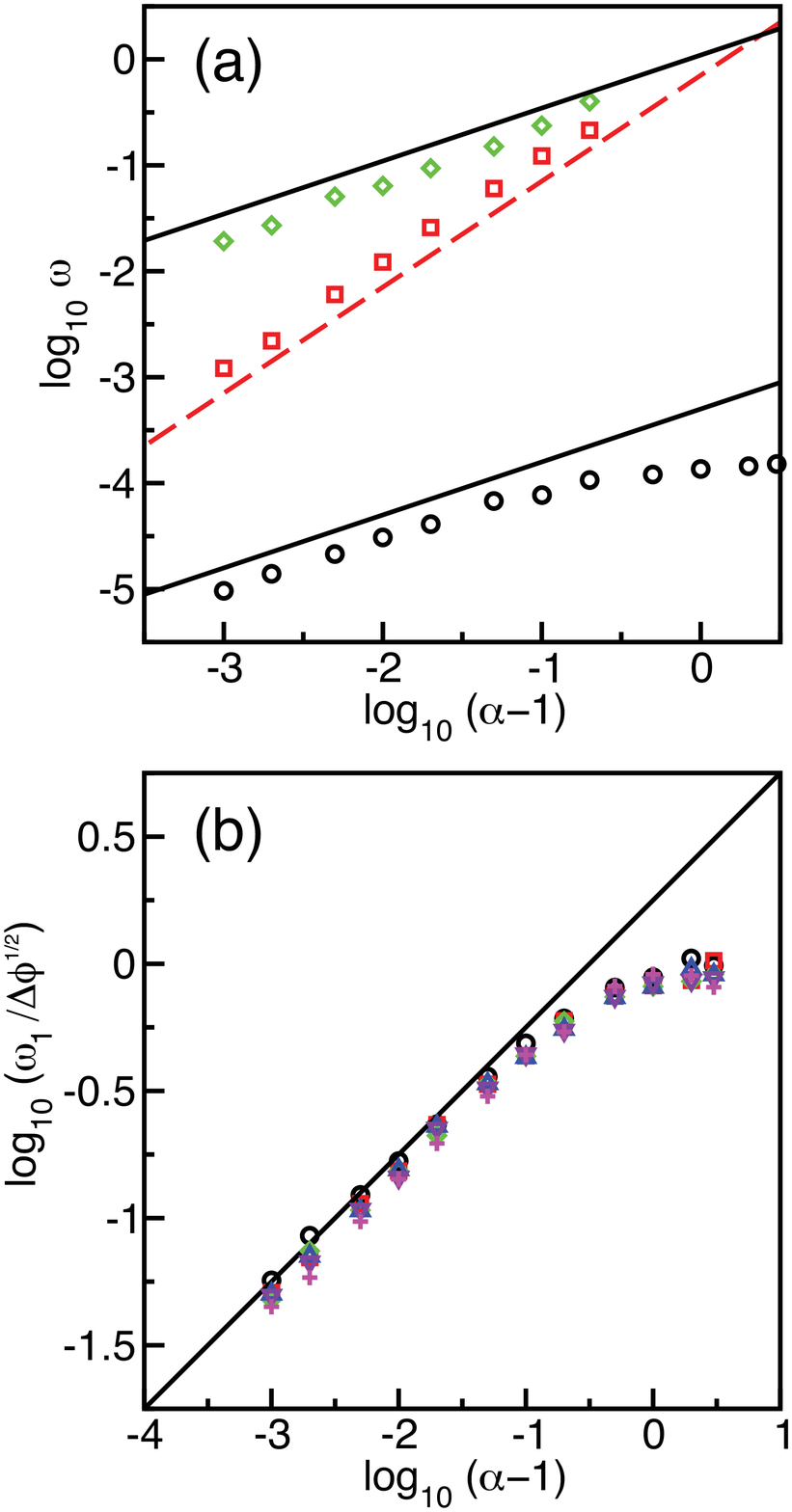}
\caption{(a) Characteristic frequencies $\omega_1$ (circles),
$\omega_2$ (squares), and $\omega_3$ (diamonds) from $D(\omega)$ as a
function of aspect ratio $\alpha - 1$ for $N=240$ ellipses in 2D at
$\Delta\phi=10^{-8}$. The solid (dashed) lines have slope $1/2$
($1)$. (b) $\omega_1/(\Delta \phi)^{1/2}$ for systems with $N=240$ ellipses in
2D at $\Delta\phi=10^{-7}$ (circles), $10^{-6}$ (squares), $10^{-5}$
(diamonds), $10^{-4}$ (upward triangles), $10^{-3}$, (downward
triangles), and $10^{-2}$ (crosses). The solid line has slope
$1/2$.}
\label{fig_char_freq}
\end{figure}

To monitor the key features of $D(\omega)$ as a function of $\Delta
\phi$ and $\alpha$, we define three characteristic frequencies as
shown in the inset to Fig.~\ref{fig_DM_2D} (b). $\omega_1$ and
$\omega_2$ identify the locations of the small and intermediate
frequency peaks in $D(\omega)$, and $\omega_3$ marks the onset of the
high-frequency plateau regime in $D(\omega)$.  For our analysis, we
define $\omega_3$ as the largest frequency ($<1$) with $D(\omega) <
0.15$, which is approximately half of the height of the plateau in
$D(\omega)$ at large frequencies.  All three characteristic
frequencies increase with aspect ratio.  Note that we only track
$\omega_2$ and $\omega_3$ for aspect ratios where $\omega_2 <
\omega_3$. For example, the intermediate and high-frequency bands 
characterized by $\omega_2$ and $\omega_3$ merge for $\alpha \ge 1.2$.   

As shown in Fig.~\ref{fig_DM_3D}, $D(\omega)$ for 3D prolate
ellipsoids displays similar behavior to that for ellipses in 2D
(Fig.~\ref{fig_DM_2D}) for aspect ratios $\alpha < 1.5$. For example,
$D(\omega)$ for ellipsoids possesses low, intermediate, and high
frequency regimes, whose characteristic frequencies $\omega_1$,
$\omega_2$, and $\omega_3$ increase with aspect ratio.  Note that the
intermediate and high-frequency bands $\omega_2$ and $\omega_3$ merge
for $\alpha>1.02$, which occurs at lower aspect ratio than the merging
of the bands in 2D.  Another significant difference is that in 3D
$D(\omega)$ extends to higher frequencies at large aspect ratios
($\alpha \gtrsim 1.2$) than $D(\omega)$ for ellipses. 

\begin{figure}
\includegraphics[width=0.4\textwidth]{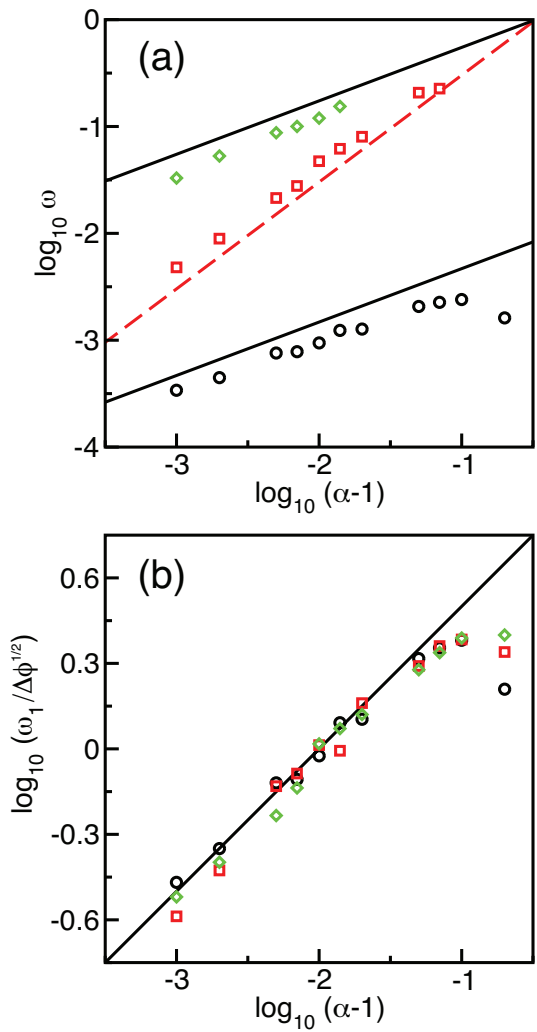}
\caption{(a) Characteristic frequencies $\omega_1$ (circles),
$\omega_2$ (squares), and $\omega_3$ (diamonds) from $D(\omega)$ as a
function of aspect ratio $\alpha - 1$ for $N=240$ prolate ellipsoids
in 3D at $\Delta\phi=10^{-6}$. The solid (dashed) lines have slope
$1/2$ ($1)$. (b) $\omega_1/(\Delta \phi)^{1/2}$ for systems with
$N=240$ prolate ellipsoids at $\Delta \phi = 10^{-6}$ (circles),
$10^{-5}$ (squares), and $10^{-4}$ (diamonds). The solid line has
slope $1/2$.}
\label{fig_char_freq_3D}
\end{figure}

We note the qualitative similarity between the $D(\omega)$ for
$\alpha=1.005$ ellipsoids shown in Fig.~\ref{fig_DM_3D} (b) and
$D(\omega)$ for $\alpha=0.96$ presented in Fig. 1 (c) of
Ref.~\cite{zeravcic} for $\omega>10^{-2}$. However, 
Zeravcic, {\it et al.} suggest that there is no weight in $D(\omega)$
for $\omega<10^{-2}$ except at $\omega=0$ for both oblate and prolate
ellipsoids, in contrast to our results in Fig.~\ref{fig_DM_3D}.

In Fig.~\ref{fig_DM_2D_comp}, we show the behavior of $D(\omega)$ for
ellipse packings as a function of compression $\Delta \phi$ for two
aspect ratios, $\alpha=1.05$ and $2$.  We find that the low-frequency band
(characterized by $\omega_1)$ depends on $\Delta \phi$, while the
intermediate and high frequency bands do not.  The intermediate and
high frequencies bands do not change significantly until the
low-frequency band centered at $\omega_1$ merges with them at
$\Delta\phi\approx 10^{-3}$ and $\approx 10^{-4}$ for $\alpha=1.05$
and $2$, respectively.

We plot the characteristic frequencies $\omega_1$, $\omega_2$, and
$\omega_3$ versus aspect ratio $\alpha-1$ for ellipse packings in
Fig.~\ref{fig_char_freq} and ellipsoid packings in
Fig.~\ref{fig_char_freq_3D}.  The characteristic frequencies obey the
following scaling laws over at least two orders of magnitude in
$\alpha-1$ and five orders of magnitude in $\Delta \phi$:
\begin{eqnarray}
\label{scaling1}
\omega_1 & \sim & (\Delta \phi)^{1/2} (\alpha-1)^{1/2}, \\
\label{scaling2}
\omega_2 & \sim & (\alpha-1),\\
\label{scaling3}
\omega_3 & \sim & (\alpha-1)^{1/2}.
\end{eqnarray}
Similar results for the scaling of $\omega_2$ and $\omega_3$ with
$\alpha-1$ were found in Ref.~\cite{zeravcic}.  We will refer to the
modes in the low-frequency band in $D(\omega)$ (with
characteristic frequency $\omega_1$) as `quartic modes', and these
will be discussed in detail Sec.~\ref{quartic_modes}. The scaling of
the quartic mode frequencies with compression, $\omega_1 \sim (\Delta
\phi)^{1/2}$, has important consequences for the linear response
behavior of ellipsoidal particles to applied stress~\cite{Mailman}.

\begin{figure}
\includegraphics[width=0.4\textwidth]{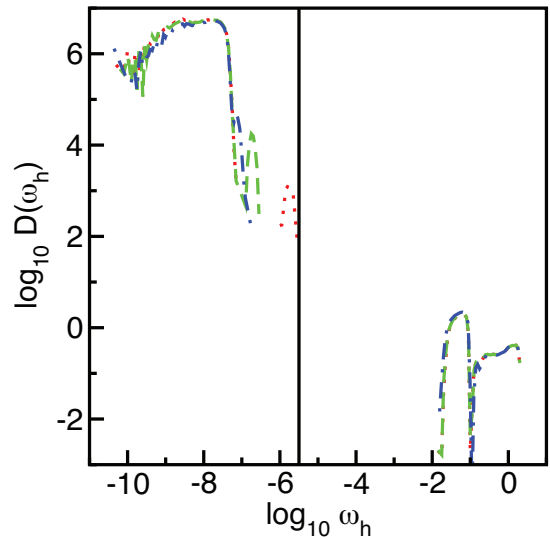}
\vspace{-0.1in}
\caption{The distribution of frequencies $D(\omega_h)$ associated with
the eigenvalues of the stiffness matrix $H$ for $N=240$ ellipse
packings as a function of compression $\Delta\phi=10^{-5}$ (dotted),
$10^{-3}$ (dashed), and $10^{-2}$ (dot-dashed) for $\alpha=1.05$. The
vertical solid line indicates the `zero-frequency' tolerance
$\omega_{\rm tol}$, which is the lowest frequency obtained for the
dynamical matrix for packings at $\alpha = 1.05$ and the smallest
compression ($\Delta \phi = 10^{-8}$) in Fig.~\ref{fig_DM_2D}.}
\label{fig_DM_stiffness_comp}
\end{figure}

\subsection{Dynamical Matrix Decomposition}
\label{decomposition}

As shown in Fig.~\ref{z}, static packings of ellipsoidal particles can
possess $z < z_{\rm iso}$ over a wide range of aspect ratio, yet as
described in Sec.~\ref{DOS}, the dynamical matrix $M$ contains a
complete spectrum of $Nd_{f} - d$ nonzero eigenvalues $m_i$ near
jamming. To investigate this intriguing property, we first calculate
the eigenvalues of the stiffness matrix $H$, show that it possesses 
$N_z$ `zero'-frequency modes whose number matches the deviation in the
contact number from the isostatic value, and then identify the separate
contributions from the stiffness and stress matrices to the dynamical
matrix eigenvalues.
  
In Fig.~\ref{fig_DM_stiffness_comp}, we show the distribution of
frequencies $D(\omega_h)$ associated with the eigenvalues of the
stiffness matrix for ellipse packings at $\alpha=1.05$ as a function
of compression $\Delta \phi$.  We find three striking features in
Fig.~\ref{fig_DM_stiffness_comp}: 1) Many modes of the stiffness
matrix exist near and below the zero-frequency threshold (determined
by the vibrational frequencies of the dynamical matrix at $\alpha = 1.05$
and $\Delta \phi=10^{-8}$), 2) Frequencies that correspond to the
low-frequency band characterized by $\omega_1$ are absent, and 3) The
nonzero frequency modes (with $\omega_h > 10^{-2}$) do not scale with
$\Delta \phi$ as pointed out for the dynamical matrix eigenvalues in
Eqs.~\ref{scaling2} and \ref{scaling3}.  Further, we find that the
number of zero-frequency modes $N_z$ of the stiffness matrix matches
the deviation in the number of contacts from the isostatic value
($N_{\rm iso} - N_c$) for each $\Delta \phi$ and aspect
ratio. Specifically, $N_z=N_{\rm iso}-N_c$ over the full range of
$\Delta \phi$ for $99.96\%$ of the more than $10^3$ packings for
aspect ratio $\alpha<1.1$ and for $100\%$ of the more than $10^3$
packings for $\alpha\ge 1.1$.

\begin{figure}
\includegraphics[width=0.4\textwidth]{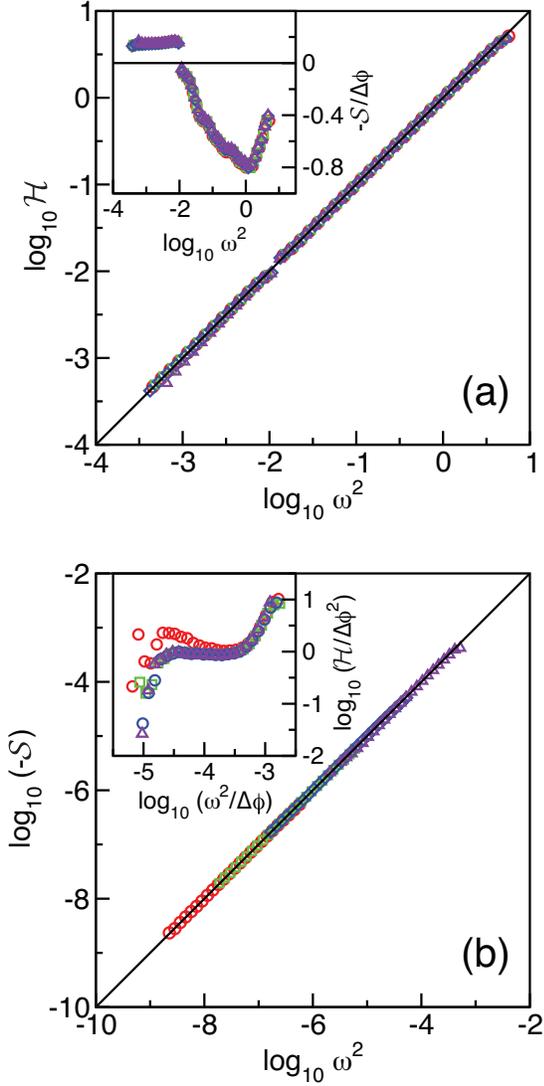}
\vspace{-0.1in}
\caption{The two contributions to the dynamical matrix eigenvalues,
(a) ${\cal H}$ and (b) $-{\cal S}$, plotted versus $\omega^2={\cal
H}-{\cal S}$ for ellipse packings with $\alpha=1.05$ and $\Delta \phi
= 10^{-6}$ (circles), $10^{-5}$ (squares), $10^{-4}$ (diamonds), and
$10^{-3}$ (triangles).  In (a) and (b), the solid lines correspond to
${\cal H}=\omega^2$ and ${\cal S}=\omega^2$.  In the main panel and
inset of (a), only modes corresponding the intermediate and high
frequency bands are included. In the main panel and inset of (b), only
modes corresponding the low frequency band are included.  The insets
to (a) and (b), which plot $-{\cal S}/(\Delta\phi)^2$ versus $\omega^2$
and ${\cal H}/(\Delta \phi)^2$ versus $\omega^2/\Delta \phi$, show the
deviations $\omega^2-{\cal H}=-{\cal S}\propto\Delta\phi$ for 
high- and intermediate-frequency modes and $\omega^2-(-{\cal
S})={\cal H}\propto(\Delta\phi)^2$ for low-frequency modes.}
\label{stiffness_matrix}
\end{figure}

In Fig.~\ref{stiffness_matrix}, we calculate the projection of the
dynamical matrix eigenvectors ${\hat m}_i$ onto the stiffness and
stress matrices, ${\cal H} = {\hat m}_i^{\dagger} H {\hat m}_i$ and
${\cal S} = {\hat m}_i^{\dagger} S {\hat m}_i$, where ${\hat
m}_i^{\dagger}$ is the transpose of ${\hat m}_i$ and $\omega_i^2 =
{\hat m}_i^{\dagger} M {\hat m}_i = {\cal H} - {\cal S}$.
Fig.~\ref{stiffness_matrix} (a) shows that for large eigenvalues
$\omega_i^2$ of the dynamical matrix ({\it i.e.} within the
intermediate and high frequency bands characterized by $\omega_2$ and
$\omega_3$ in Fig.~\ref{fig_DM_2D}), the eigenvalues of the stiffness
and dynamical matrices are approximately the same, ${\cal H} \approx
\omega_i^2$. The deviation $\omega_i^2 - {\cal H}= -{\cal S}$, shown
in the inset to Fig.~\ref{stiffness_matrix} (a), scales linearly with
$\Delta\phi$.  Thus, we find that the intermediate and high frequency
modes for packings of ellipsoidal particles are stabilized by the
stiffness matrix $H$.

\begin{figure}
\includegraphics[width=0.4\textwidth]{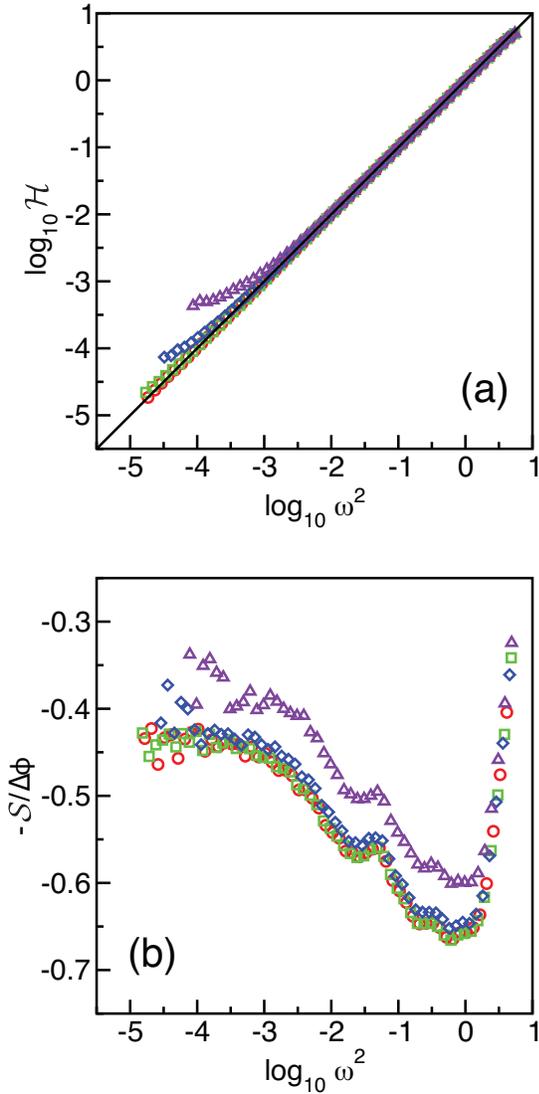}
\vspace{-0.1in}
\caption{The two contributions to the dynamical matrix eigenvalues,
(a) ${\cal H}$ and (b) $-{\cal S}/\Delta \phi$, plotted versus the
eigenvalues of the dynamical matrix $\omega^2$ for bidisperse disk
packings at $\Delta \phi = 10^{-6}$ (circles), $10^{-5}$ (squares),
$10^{-4}$ (diamonds), and $10^{-3}$ (triangles).  In (a) the solid
line corresponds to ${\cal H}=\omega^2$. Note that $-{\cal S} < 0$
over the entire range of frequencies.}
\label{decompose2d}
\end{figure}

In the main panel of Fig.~\ref{stiffness_matrix} (b), we show that for
frequencies in the lowest frequency band (characterized by $\omega_1$)
the eigenvalues of the stress and dynamical matrices are approximately
the same, $-{\cal S} \approx \omega_i^2$.  In the inset to
Fig.~\ref{stiffness_matrix} (b), we show that the deviation
$\omega_i^2 - ({-\cal S}) = {\cal H}$ scales as $(\Delta\phi)^2$.
Thus, we find that the lowest frequency modes for packings of
ellipsoidal particles are stabilized by the stress matrix $-S$ over a
wide range of compression $\Delta \phi$. Similar results were found
previously for packings of hard ellipsoidal
particles~\cite{torquato_ellipse}.  In contrast, for static packings
of spherical particles, the stress matrix contributions to the
dynamical matrix are destabilizing with $-{\cal S} < 0$ for all
frequencies near jamming, and ${\cal H}$ stabilizes the packings as 
shown in Fig.~\ref{decompose2d}.

\subsection{Quartic modes}
\label{quartic_modes}

We showed in Sec.~\ref{DOS} that the dynamical matrix $M$ for packings
of ellipsoidal particles contains a complete spectrum of $Nd_{f}
- d$ nonzero eigenvalues $m_i$ for $\Delta \phi >0$ despite that fact
that $z < z_{\rm iso}$. Further, we showed that the modes in the lowest
frequency band scale as $\omega_1 \sim (\Delta \phi)^{1/2}$ in the
$\Delta \phi \rightarrow 0$ limit.  What happens at jamming onset when
$\Delta \phi = 0$, {\it i.e.} are these low-frequency modes that become 
true zero-frequency modes at $\Delta \phi=0$ stabilized or
destabilized by higher-order terms in the expansion of the potential energy in
powers of the perturbation amplitude?

\begin{figure}[h]
\begin{center}
\scalebox{0.5}{\includegraphics{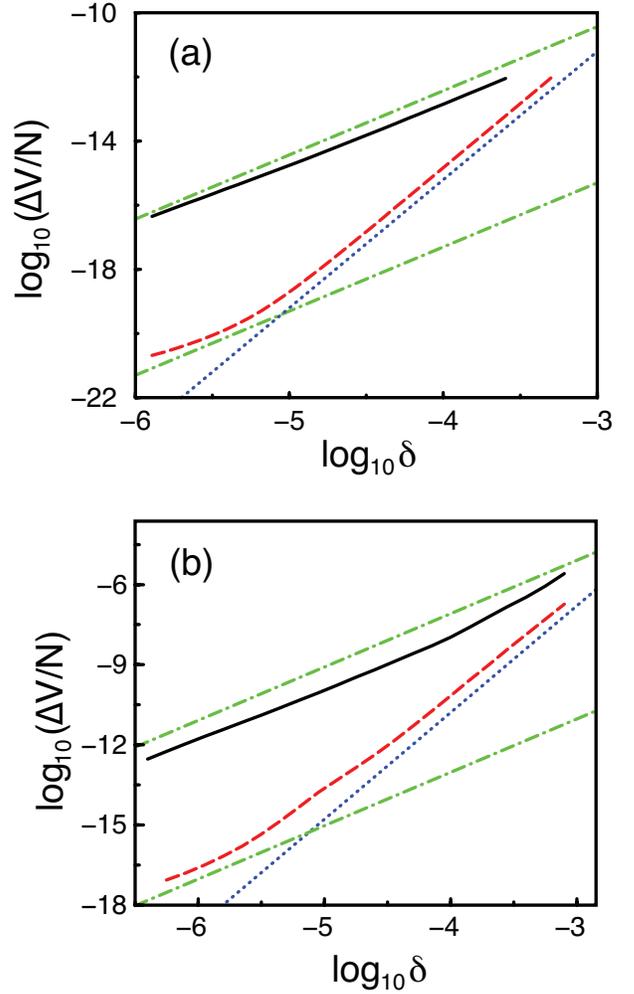}}
\caption{The change in the total potential energy $\Delta V$
(normalized by the particle number $N$) before and after applying the
perturbation in Eq.~\ref{perturbation} with amplitude $\delta$ along
the eigenvector that corresponds to the lowest nontrivial eigenvalue
of the dynamical matrix for packings of (a) disks (solid line) and
ellipses with $\alpha=1.1$ (dashed line) and (b) spheres (solid line)
and prolate ellipsoids with $\alpha=1.1$ (dashed line) at $\Delta
\phi=10^{-7}$.  The dot-dashed (dotted) lines have slope $2$ ($4$).}
\label{fig7b}
\end{center}
\end{figure}

To investigate this question, we apply the following deformation 
to static packings of ellipsoidal particles: 
\begin{equation}
\label{perturbation}
{\vec u} = {\vec u}_0 + \delta {\hat m}_i,
\end{equation}
where $\delta$ is the amplitude of the perturbation, ${\hat m}_i$ is
an eigenvector of the dynamical matrix, and ${\vec u}_0$ is the point
in configuration space corresponding to the original static packing,
followed by conjugate gradient energy minimization.  We then measure
the change in the total potential energy before and after the
perturbation, $\Delta V$.
 
\begin{figure}[h]
\begin{center}
\scalebox{0.4}{\includegraphics{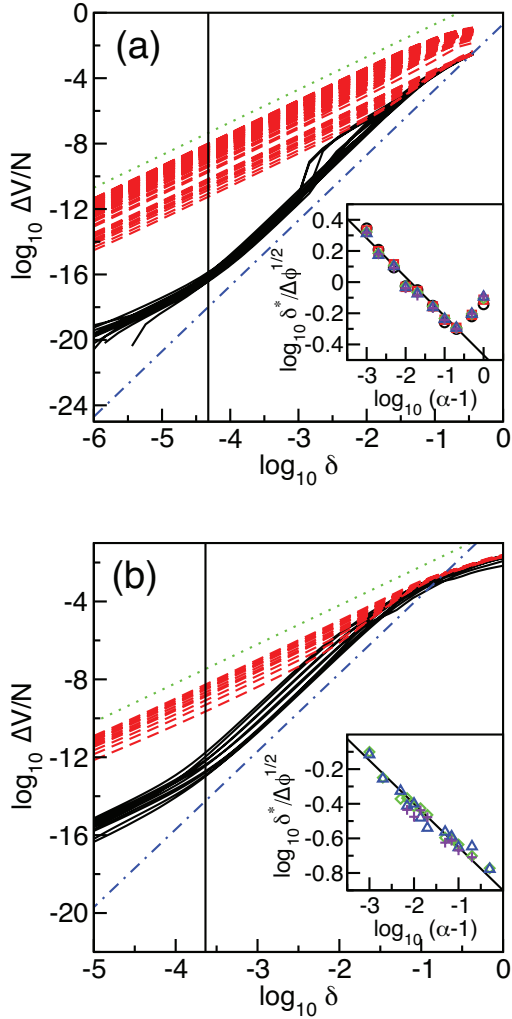}}
\caption{The change in the total potential energy $\Delta V/N$ for
perturbations along the `quartic' modes (solid) and all other modes
(dashed) as a function of amplitude $\delta$ for (a) ellipses and (b)
prolate ellipsoids with $\alpha=1.1$ for $\Delta \phi = 10^{-7}$. The
dotted (dot-dashed) lines have slope $2$ ($4$). The solid vertical
lines indicate the characteristic amplitude $\delta^*$ at which
$\Delta V/N$ crosses over from quadratic to quartic scaling averaged
over the quartic modes. The insets show the scaling of 
$\delta^*/(\Delta \phi)^{1/2}$ with $\alpha-1$ for several
values of compression: $\Delta \phi = 10^{-8}$ (circles) and $10^{-7}$
(squares) for 2D systems and $10^{-6}$ (diamonds), $10^{-5}$
(triangles), and $10^{-4}$ (pluses) for both 2D and 3D systems. The
solid lines in the insets have slope $-0.25$.}
\label{fig15}
\end{center}
\end{figure}

We plot $\Delta V/N$ versus $\delta$ in Fig.~\ref{fig7b} for
perturbations along eigenvectors that correspond to the smallest
nontrivial eigenvalue $m_1 = \omega_{\rm min}^2$ of the dynamical
matrix for static packings of (a) disks and ellipses and (b) spheres
and prolate ellipsoids at $\Delta \phi = 10^{-7}$. As expected, for
disks and spheres, we find that $\Delta V/N \approx m \omega_{\rm
min}^2 \delta^2$ over a wide range of $\delta$ in response to
perturbations along eigenvectors that correspond to the smallest
nontrivial eigenvalue. In contrast, we find novel behavior for $\Delta
V/N$ when we apply perturbations along the eigendirection that
corresponds the lowest nonzero eigenvalue of the dynamical matrix for
packings of ellipsoidal particles.  In Fig.~\ref{fig15}, we show that
$\Delta V/N$ obeys
\begin{equation}
\label{quartic_equation}
\frac{\Delta V}{N}=\frac{m}{2} \omega_k^2\delta^2+ c_k \delta^4,
\end{equation}
where $\omega_k \propto\Delta\phi^{1/2}$ and the constants $c_k>0$,
for perturbations along all modes $k$ in the lowest frequency band of
$D(\omega)$ for packings of ellipsoidal particles when we do not include
changes in the contact network following the perturbation and
relaxation.  (See Appendix~\ref{contact_formation} for measurements of
$\Delta V/N$ when we include changes in the contact network.)
Eigenmodes in the lowest frequency band are termed `quartic' because
at $\Delta \phi = 0$ they are stabilized by quartic terms in the
expansion of the total potential energy with respect to small
displacements~\cite{Mailman}.

For $\delta \ll \delta_k^*$, the change in potential energy scales as
$\Delta V/N\sim \omega_k^2 \delta^2$, whereas $\Delta V/N \sim c_k
\delta^4$ for $\delta \gg \delta_k^*$, where the characteristic
perturbation amplitude $\delta_k^* = \omega_k \sqrt{m/2 c_k}$.  In the
insets to Fig.~\ref{fig15} (a) and (b), we show that the characteristic
perturbation amplitude averaged over modes in the lowest frequency
band scale as $\delta^* \sim (\Delta \phi)^{1/2}/(\alpha-1)^{1/4}$ for
static packings of ellipses in 2D and prolate ellipsoids in 3D, which
indicates that the $c_k$ possess nontrivial dependence on aspect ratio
$\alpha$.

\begin{figure}
\includegraphics[width=0.4\textwidth]{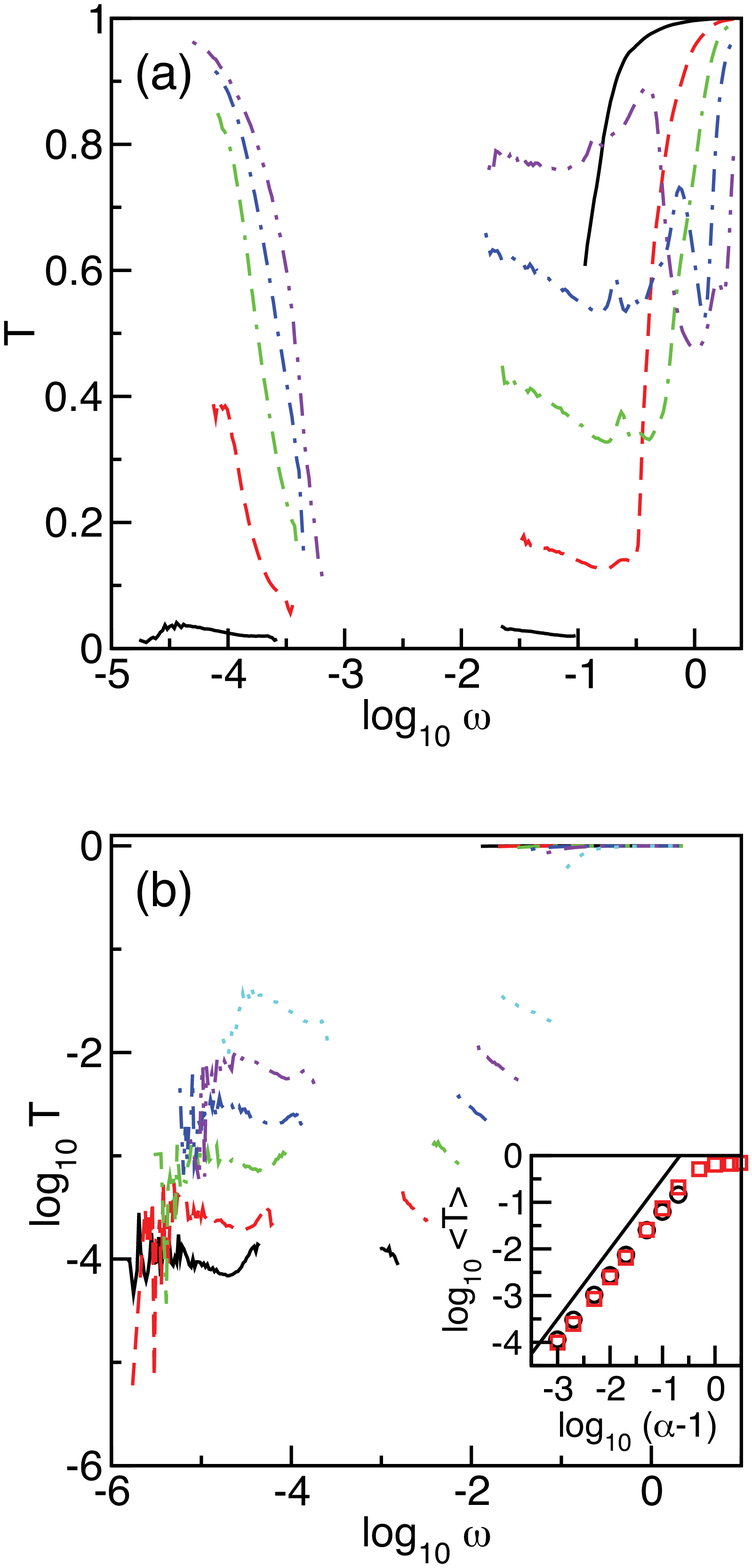}
\caption{The contribution of the translational degrees of freedom $T$
to each eigenvector ${\hat m}$ of the dynamical matrix versus
frequency $\omega$ in packings of ellipses in 2D at $\Delta
\phi=10^{-7}$.  (a) shows data for aspect ratios $\alpha=1.05$ (black
solid), $1.2$ (red dashed), $1.5$ (green dash-dash-dot), $2.0$ (blue
dash-dot), and $4.0$ (purple dot-dot-dash) and (b) shows data for
aspect ratios $\alpha=1.001$ (black solid), $1.002$ (red dashed),
$1.005$ (green dot-dot-dash), $1.01$ (blue dash-dot), $1.02$ (purple
dot-dot-dash), and $1.05$ (cyan dotted). The inset to (b) shows
$\langle T\rangle$ averaged over modes in the lowest (squares) and
intermediate (circles) frequency regimes. The solid line has slope
$1.5$.}
\label{fig_comp_2D}
\end{figure}

\begin{figure}
\includegraphics[width=0.4\textwidth]{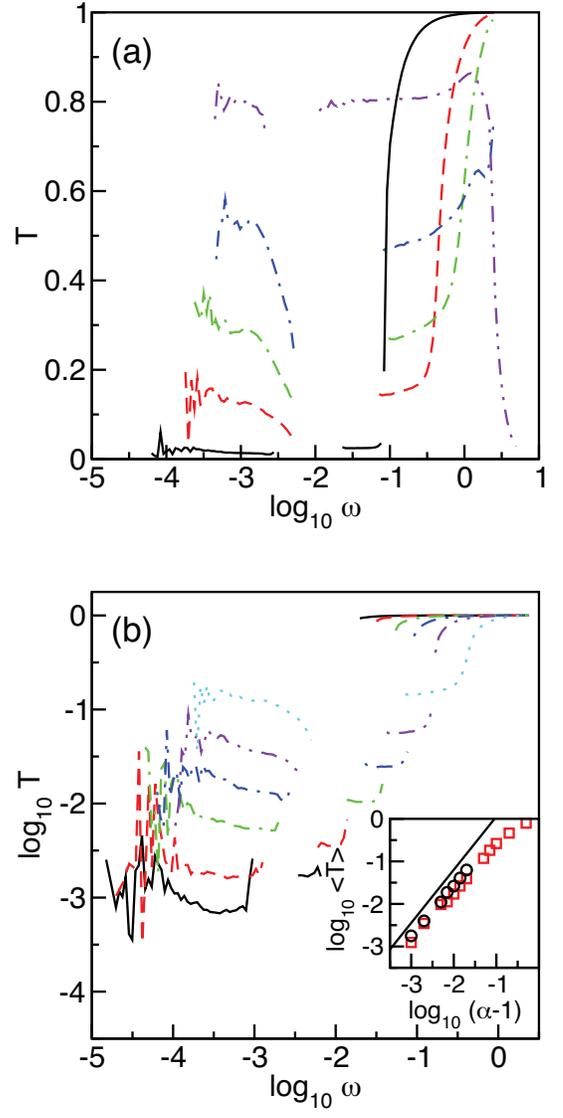}
\caption{The contribution of the translational degrees of freedom $T$
to each eigenvector ${\hat m}$ of the dynamical matrix versus
frequency $\omega$ in packings of prolate ellipsoids in 3D at $\Delta
\phi=10^{-6}$.  (a) shows data for aspect ratios $\alpha=1.01$ (black
solid), $1.05$ (red dashed), $1.1$ (green dash-dash-dot), $1.2$ (blue
dash-dot), and $1.5$ (purple dot-dot-dash) and (b) shows data for
aspect ratios $\alpha=1.001$ (black solid), $1.002$ (red dashed),
$1.005$ (green dash-dash-dot), $1.01$ (blue dash-dot), $1.02$ (purple
dot-dot-dash), and $1.05$ (cyan dotted). The inset to (b) shows $\langle
T\rangle$ averaged over modes in the lowest (squares) and intermediate
(circles) frequency regimes. The solid line has slope $1.25$.}
\label{fig_comp_3D}
\end{figure}

The quartic modes have additional interesting features.  For example,
quartic modes are dominated by the rotational rather than
translational degrees of freedom.  We identify the relative
contributions of the translational and rotational degrees of freedom
to the eigenvectors of the dynamical matrix in Figs.~\ref{fig_comp_2D}
and~\ref{fig_comp_3D}.  The contribution of the translational degrees
of freedom to eigenvector ${\hat m}_i$ is defined as
\begin{equation}
\label{ti}
T_i = \sum_{j=1}^{Nd_f} \sum_{\lambda} (m_{\lambda i}^j)^2,
\end{equation}
where the sum over $\lambda$ includes $x$ and $y$ in 2D and $x$, $y$,
and $z$ in 3D and the eigenvectors are indexed in increasing order of
the corresponding eigenvalues. Since the eigenvectors are normalized,
the rotational contribution to each eigenvector is $R_i = 1 -
T_i$. 

For both ellipses in 2D and prolate ellipsoids in 3D, we find that at
low aspect ratios ($\alpha < 1.1$), the first $N$ ($2N$) modes in 2D
(3D) are predominately rotational and the remaining $2N$ ($3N$) modes
in 2D (3D) are predominately translational. In the inset to
Figs.~\ref{fig_comp_2D}(b) and~\ref{fig_comp_3D}, we show that $T$
increases as $(\alpha - 1)^{\zeta}$, where $\zeta \approx 1.5$
($1.25$) for ellipses (prolate ellipsoids), for both the low and
intermediate frequency modes.  For $\alpha>1.2$, we find mode-mixing,
especially at intermediate frequencies, where modes have finite
contributions from both the rotational and translational degrees of
freedom. For $\alpha \le 1.2$, the modes become increasingly more
translational with increasing frequency. For $\alpha > 1.2$, the
modes become more rotational in character at the highest
frequencies. Our results show that the modes with
significant rotational content at low $\alpha$ correspond to modes in
the low and intermediate frequency bands of $D(\omega)$, while the
modes with significant translational content at low $\alpha$
correspond to modes in the high frequency band of $D(\omega)$.

\subsection{Protocol dependence}
\label{protocol}

We performed several checks to test the robustness and accuracy of our
calculations of the density of vibrational modes in the harmonic
approximation $D(\omega)$ for static packings of ellipsoidal
particles: 1) We compared $D(\omega)$ obtained from static packings of
ellipsoidal particles using Perram and Wertheim's exact expression
(Eq.~\ref{sigma_functional}) for the contact distance between pairs of
ellipsoidal particles and the Gay-Berne approximation described in
Sec.~\ref{contact}; 2) We calculated $D(\omega)$ for static packings
as a function of the tolerance used to terminate energy minimization
for both the MD and CG methods; and 3) We studied the system-size
dependence of $D(\omega)$ in systems ranging from $N=30$ to $960$
particles.

\begin{figure}[h]
\begin{center}
\scalebox{0.43}{\includegraphics{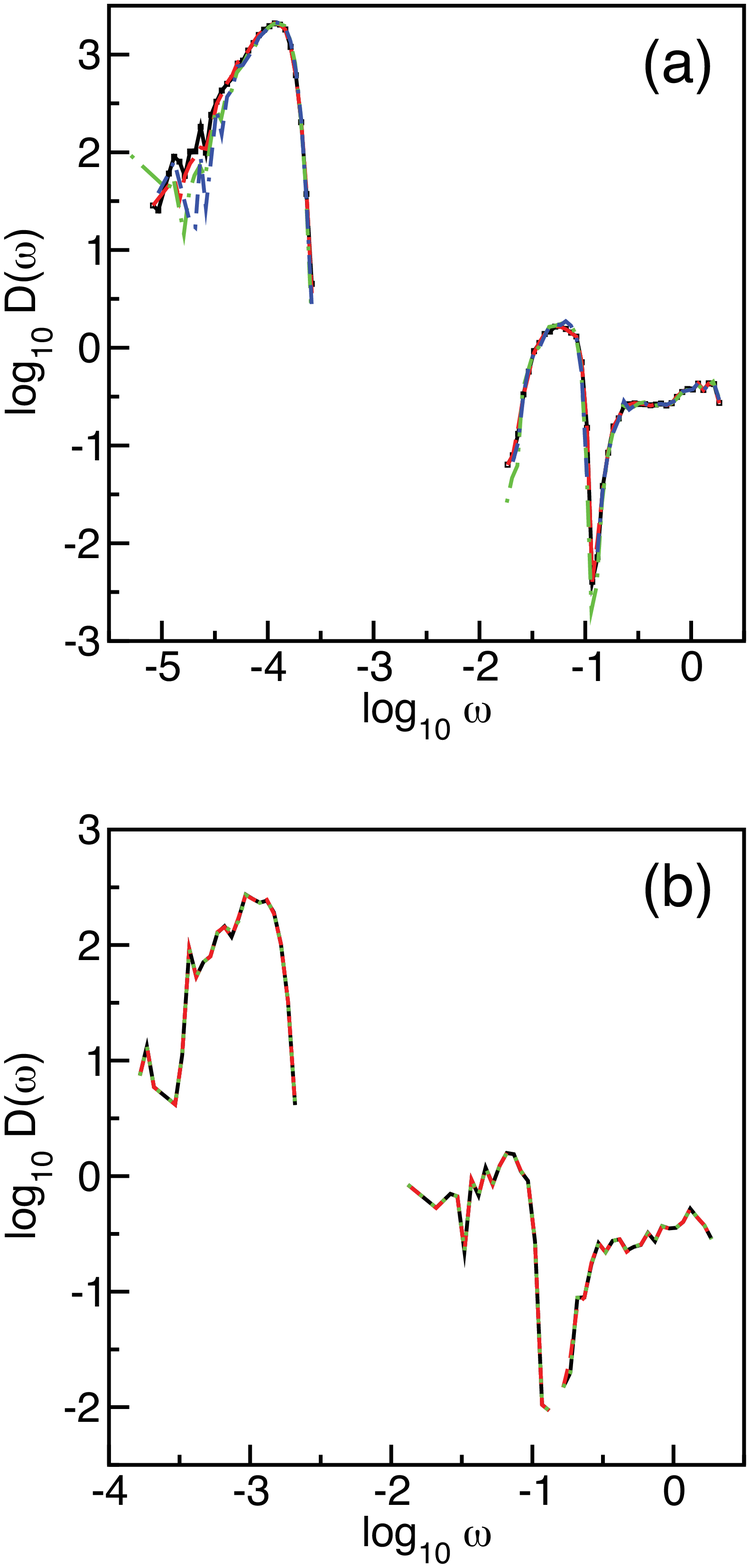}}
\caption{(a) Density of vibrational modes in the harmonic
approximation $D(\omega)$ for $N=30$ ellipses with $\alpha=1.05$ at
$\Delta \phi = 10^{-7}$ using the Perram and Wertheim exact contact
distance between pairs of ellipses with CG energy minimization
tolerance $V_{\rm tol} = 10^{-16}$ (green dot-dashed) and $V_{\rm tol}
= 10^{-8}$ (blue dash-dash-dotted) or the Gay-Berne approximation with
$V_{\rm tol} = 10^{-16}$ (black solid) and $V_{\rm tol} = 10^{-8}$
(red dashed). (b) $D(\omega)$ for $N=12$ ellipses with $\alpha=1.05$ at
$\Delta \phi = 10^{-5}$ using the Perram and Wertheim exact contact
distance with CG energy minimization tolerance $V_{\rm tol} =
10^{-16}$ (solid black), and MD energy minimization tolerance $K_{\rm
tol} = 10^{-16}$ (red dashed) and $10^{-24}$ (green dotted).}
\label{fig11}
\end{center}
\end{figure}

In Fig.~\ref{fig11}, we show that the density of vibrational modes
$D(\omega)$ is nearly the same when we use the Perram and Wertheim
exact expression and the Gay-Berne approximation to the contact
distance for ellipse-shaped particles.  $D(\omega)$ for static
packings of ellipse-shaped particles is also not dependent on $V_{\rm
tol}$, which controls the accuracy of the conjugate gradient energy
minimization (Sec.~\ref{packing}), for sufficiently small values. Our
calculations in Fig.~\ref{fig11} (b) also show that $D(\omega)$ is not
sensitive to the energy minimization procedure ({\it i.e.} MD vs. CG)
for small values of the minimization tolerance $K_{\rm tol}$.

\begin{figure}[h]
\begin{center}
\scalebox{0.45}{\includegraphics{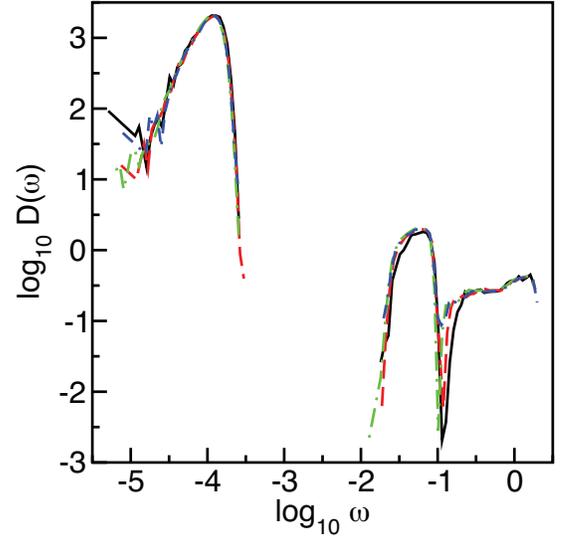}}
\caption{Density of vibrational modes in the harmonic approximation
$D(\omega)$ for ellipses in 2D with aspect ratio $\alpha=1.05$,
$\Delta \phi = 10^{-7}$, and system size $N=30$ (black solid), $120$
(red dashed), $240$ (green dot-dashed), and $960$
(blue dash-dash-dotted).}
\label{system_size}
\end{center}
\end{figure}

In addition, key features of the density of vibrational modes are not
strongly dependent on system size.  For example, in
Fig.~\ref{system_size}, we show $D(\omega)$ for ellipses in 2D at
aspect ratio $\alpha = 1.05$ and compression $\Delta \phi = 10^{-7}$
over a range of system sizes from $N=30$ to $960$. (For reference,
$D(\omega)$ at fixed system size $N=240$ and $\Delta \phi = 10^{-8}$
over a range of aspect ratios is shown in Fig.~\ref{fig_DM_2D}.)
$D(\omega)$ in the low and intermediate frequency bands and plateau
region overlap for all system sizes.  The only feature of $D(\omega)$
that changes with system size is that successively lower frequency,
long wavelength translational modes extend from the plateau region as
system size increases.  In the large system-size limit $N > (\phi
-\phi_J)^{-2}$, which we do not reach in these studies, the lowest
frequency modes will scale as $D(\omega) \sim \omega^{d-1}$.

\section{Conclusions}
\label{conclusion}

We performed extensive numerical simulations of static packings of
frictionless, purely repulsive ellipses in 2D and prolate ellipsoids
in 3D as a function of aspect ratio $\alpha$ and compression from
jamming onset $\Delta \phi$.  We found several important results that
highlight the significant differences between amorphous
packings of spherical and ellipsoidal particles near jamming.  First,
as found previously, static packings of ellipsoidal particles
generically satisfy $z < z_{\rm
iso}$~\cite{Mailman,zeravcic,torquato_ellipse,delaney}; {\it i.e.}
they possess fewer contacts than the minimum required for mechanical
stability as predicted by counting arguments that assume all contacts
give rise to linearly independent constraints on particle
positions. Second, we decomposed the dynamical matrix $M=H-S$ into the
stiffness $H$ and stress $S$ matrices. We found that the stiffness
matrix possesses $N(z_{\rm iso} - z)$ eigenmodes ${\hat e}_0$ with
zero eigenvalues over a wide range of compressions $(\Delta \phi >
0)$. Third, we found that the modes ${\hat e}_0$ are nearly
eigenvectors of the dynamical matrix (and the stress matrix $-S$) with
eigenvalues that scale as $c \Delta \phi$, with $c>0$, and thus finite
compression stabilizes packings of ellipsoidal
particles~\cite{Mailman}.  At jamming onset, the harmonic response of
packings of ellipsoidal particles vanishes, and the total potential
energy scales as $\delta^4$ for perturbations by amplitude $\delta$
along these `quartic' modes, ${\hat e}_0$.  In addition, we have shown
that these results are robust; for example, the density of vibrational
modes $D(\omega)$ (in the harmonic approximation) is not sensitive to
the error tolerance of the energy minimization procedure, the system
size, and the accuracy of the determination of the interparticle contacts
over the range of parameters employed in the simulations.

These results raise several fundamental questions for static
granular packings: 1) Which classes of particle shapes give rise to
quartic modes?; 2) Is there a more general isostatic counting argument
that can predict the number of quartic modes at jamming onset (for a
given packing-generation protocol)?; and 3) Are systems with quartic
modes even more anharmonic~\cite{schreck2011} than packings of
spherical particles in the presence of thermal and other sources of
fluctuations?  We will address these important questions in our future
studies. 

\begin{acknowledgments}
Support from NSF grant numbers DMR-0905880 (BC and MM) and DMS-0835742
(CS and CO) is acknowledged.  We also thank T. Bertrand, M. Bi, and
M. Shattuck for helpful discussions.
\end{acknowledgments}

\appendix 

\section{Scaling Behavior of the Total Potential Energy}
\label{contact_formation}

\begin{figure}[h]
\begin{center}
\scalebox{0.4}{\includegraphics{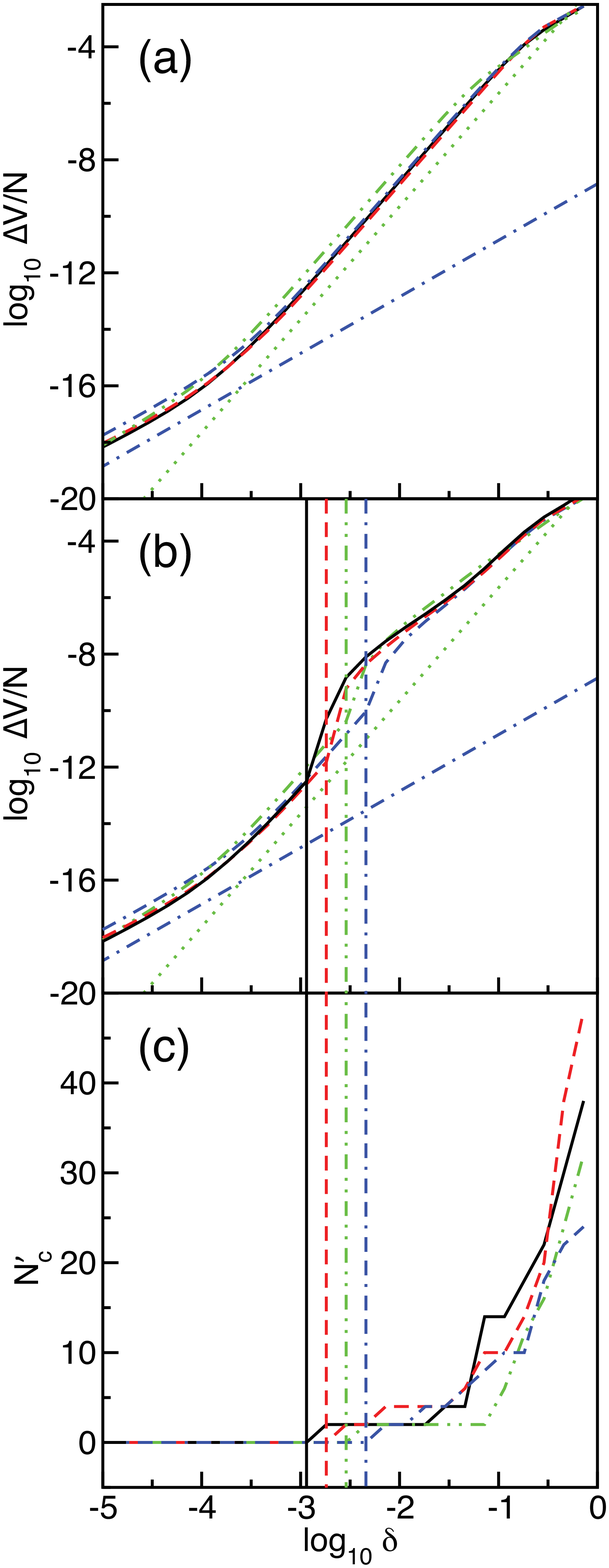}}
\caption{Change in the total potential energy $\Delta V/N$ when we (a)
do not allow the system to gain contacts or (b) allow the system to
gain contacts versus the amplitude of the perturbation $\delta$ along
several `quartic' modes (mode $17$: black solid, mode $23$: red
dashed, mode $38$: green dot-dot-dashed, and mode $78$: blue
dash-dash-dotted) from a static packing of $N=240$ ellipse-shaped
particles at $\Delta \phi=10^{-8}$ and $\alpha = 1.002$.  The dotted
(dot-dashed) line has slope $4$ ($2$). (c) The number of new contacts
$N_c'$ that differ from the original contact network as a function of
$\delta$ for each mode in (a) and (b). The vertical lines indicate the
$\delta$ at which the first new contact forms for each mode.}
\label{contact_breaking_fig}
\end{center}
\end{figure}

The scaling behavior of $\Delta V/N$ (shown in Figs.~\ref{fig7b}
and~\ref{fig15}) as a function of the amplitude $\delta$ of the
perturbation along the eigenmodes of the dynamical matrix is valid
only when the original contact network of the perturbed static packing
does not change.  As shown in Fig.~\ref{contact_breaking_fig}, $\Delta
V/N$ does not obey the power-law scaling described in
Eq.~\ref{quartic_equation} when new interparticle contacts form.  Note
that changes in the contact network are more likely for systems with
$\alpha \sim 1$ as shown previously in Ref.~\cite{schreck2011}.  In a
future publication, we will measure the critical perturbation amplitude
$\delta^k_c$ below which new contacts do not form and existing
contacts do not change for each mode $k$.  This work is closely related 
to determining the nonlinear vibrational response of packings of 
ellipsoidal and other anisotropic particles.

\section{Dynamical Matrix for Ellipse-shaped Particles}
\label{appendixb} 

In this Appendix, we provide explicit expressions for the dynamical
matrix elements (Eq.~\ref{eq:DM_eq}) for ellipse-shaped particles that
interact via the purely repulsive linear spring potential
(Eq.~\ref{eq:ellipse_energy}).  The nine dynamical matrix elements for
$i\ne j$ are
\begin{eqnarray}
M_{x_ix_j}&=&\frac{F_{ij}}{r_{ij}^3}y_{ij}^2\bigg(1
-\frac{\partial^2 \ln \sigma_{ij} }{\partial\psi_{ij}^2}
+\bigg(\frac{\partial \ln \sigma_{ij} }{\partial\psi_{ij}} \bigg)^2\bigg)\nonumber+\\
&&\frac{G_{ij}}{r_{ij}^2}\bigg(x_{ij}+y_{ij}\frac{\partial 
\ln \sigma_{ij} }{\partial\psi_{ij}}\bigg)^2\\
M_{x_iy_j}&=&\frac{F_{ij}}{r_{ij}^3}x_{ij}y_{ij}\bigg(1
-\frac{\partial^2 \ln \sigma_{ij} }{\partial\psi_{ij}^2}
+\bigg(\frac{\partial \ln \sigma_{ij} }{\partial\psi_{ij}} \bigg)^2\bigg)\nonumber+\\
&&\frac{G_{ij}}{r_{ij}^2}\bigg(x_{ij}+y_{ij}\frac{\partial 
\ln \sigma_{ij} }{\partial\psi_{ij}}\bigg)
\bigg(y_{ij}-x_{ij}\frac{\partial \ln \sigma_{ij} }{\partial
\psi_{ij}}\bigg)\nonumber\\\\
M_{x_i\theta_j}&=& l^{-1} \bigg(\frac{F_{ij}}{r_{ij}} \bigg(y_{ij}
\frac{\partial^2 \ln \sigma_{ij} }{\partial\psi_{ij}\partial \theta_{j}}-\nonumber\\
&&\frac{\partial \ln \sigma_{ij} }{\partial\theta_{j}}\bigg(x_{ij} +y_{ij}
\frac{\partial \ln \sigma_{ij} }{\partial\psi_{ij}}\bigg)\bigg)+ \nonumber\\
&&G_{ij}\frac{\partial \ln \sigma_{ij} }{\partial\theta_{j}}
\bigg(x_{ij}+y_{ij}\frac{\partial \ln \sigma_{ij} }{\partial
\psi_{ij}}\bigg)\bigg)\bigg)\\
M_{y_ix_j}&=&M_{x_iy_j}\\
M_{y_iy_j}&=&\frac{F_{ij}}{r_{ij}^3}x_{ij}^2\bigg(
1-\frac{\partial^2 \ln \sigma_{ij} }{\partial\psi_{ij}^2}
+\bigg(\frac{\partial \ln \sigma_{ij} }{\partial\psi_{ij}} \bigg)^2\bigg)\nonumber+\\
&&\frac{G_{ij}}{r_{ij}^2}\bigg(y_{ij}-x_{ij}\frac{\partial 
\ln \sigma_{ij} }{\partial\psi_{ij}}\bigg)^2\\
M_{y_i\theta_j}&=& l^{-1} \bigg(\frac{F_{ij}}{r_{ij}} \bigg(x_{ij}
\frac{\partial^2 \ln \sigma_{ij} }{\partial\psi_{ij}\partial \theta_{j}}-\nonumber\\
&&\frac{\partial \ln \sigma_{ij} }{\partial\theta_{j}}\bigg(y_{ij}- x_{ij}
\frac{\partial \ln \sigma_{ij} }{\partial\psi_{ij}}\bigg)\bigg)+ \nonumber\\
&&G_{ij}\frac{\partial \ln \sigma_{ij} }{\partial\theta_{j}}
\bigg(y_{ij}-x_{ij}\frac{\partial \ln \sigma_{ij} }{\partial
\psi_{ij}}\bigg)\bigg)\bigg)\\
M_{\theta_ix_j}&=&M_{x_j\theta_i}\\\
M_{\theta_iy_j}&=&M_{y_j\theta_i}\\\
M_{\theta_i\theta_j}&=&l^{-2} \bigg(F_{ij}r_{ij}\bigg(
\frac{\partial \ln \sigma_{ij} }{\partial\theta_i}\frac{\partial 
\ln \sigma_{ij} }{\partial\theta_j}-
\frac{\partial^2 \ln \sigma_{ij} }{\partial\theta_i^2}\bigg)+\nonumber\\
& & G_{ij}r_{ij}^2\bigg(\frac{\partial \ln \sigma_{ij} }{\partial\theta_i}
\frac{\partial \ln \sigma_{ij} }{\partial\theta_j}\bigg)\bigg)
\label{eq:DM_ellipse_1}
\end{eqnarray}
and the nine dynamical matrix elements for $i=j$ are
\begin{eqnarray}
M_{x_ix_i}&=&-\sum_jM_{x_jx_i}\\
M_{x_iy_i}&=&-\sum_jM_{x_jy_i}\\
M_{x_i\theta_i}&=&-\sum_jM_{x_j\theta_i}\\
M_{y_ix_i}&=&M_{x_iy_j}\\
M_{y_iy_i}&=&-\sum_jM_{y_jy_i}\\
M_{y_i\theta_i}&=&-\sum_jM_{y_j\theta_i}\\
M_{\theta_ix_i}&=&M_{x_i\theta_i}\\
M_{\theta_iy_i}&=&M_{y_i\theta_i}\\
M_{\theta_i\theta_i}&=& l^{-2} \bigg(F_{ij}r_{ij}\bigg(
\bigg(\frac{\partial \ln \sigma_{ij} }{\partial\theta_i}\bigg)^2-
\frac{\partial^2 \ln \sigma_{ij} }{\partial\theta_i^2}\bigg)+\nonumber\\
&&G_{ij}r_{ij}^2\bigg(\frac{\partial \ln \sigma_{ij} }{\partial \theta_i}\bigg)^2\bigg),
\label{eq:DM_ellipse_2}
\end{eqnarray}
%%%%%%%%%%%%
where $\psi_{ij}$ is the polar angle defined in Fig.~\ref{torque_schematic}, $l=\sqrt{I/m}$, $F_{ij}$ is
given in Eq.~\ref{fij},
%%%%%%%%%%%%
\begin{eqnarray}
G_{ij}&=&\left| \frac{\partial^2 V_{ij}(r_{ij}/\sigma_{ij})}{\partial 
r_{ij}^2}\right|=\sigma_{ij}^{-2},\\
%%%%%%%%%%%%
\frac{\partial \ln \sigma_{ij} }{\partial\theta_{i}}&=&\frac{\chi}{2}\bigg(\frac{\sigma_{ij}}{\sigma_{ij}^0}\bigg)^2
(\eta_++\eta_-)\bigg(\beta\cos^2(\theta_i-\psi_{ij})+\nonumber\\
&&\frac{\chi}{2}(\eta_+-\eta_-)\cos^2(\theta_i-\theta_j)\bigg),\\
%%%%%%%%%%%%
\frac{\partial^2 \ln \sigma_{ij} }{\partial\psi_{ij}^2}&=&\frac{\chi}{2}\bigg(\frac{\sigma_{ij}}{\sigma_{ij}^0}\bigg)^2
\bigg(\big(1+\chi\cos(\theta_i-\theta_j)\big)\big(\nu_+^2-\eta_+^2\big)\nonumber\\
&&+\big(1-\chi\cos(\theta_i-\theta_j)\big)\big(\nu_-^2-\eta_-^2\big)\bigg)\nonumber\\
&&+2\bigg(\frac{\partial \ln \sigma_{ij} }{\partial\psi_{ij}}\bigg)^2+
\frac{\partial^2 \ln \sigma_{ij} }{\partial\psi_{ij}^2}\bigg|_{\rm corr},\\
%%%%%%%%%%%%
\frac{\partial^2 \ln \sigma_{ij} }{\partial\psi_{ij}\partial \theta_{i}}&=&\frac{\chi}{2}
\bigg(\frac{\sigma_{ij}}{\sigma_{ij}^0}\bigg)^2
(\beta^{-1}\sin(\theta_i-\psi_{ij})(\eta_+-\eta_-)-\nonumber\\
&&\beta^{-1}\sin(\theta_i-\psi_{ij})\nonumber\\
&&(\eta_+'-\eta_-')+\nonumber\\
&&\chi\sin(\theta_i-\theta_j)(\eta_+\eta_+'-\eta_-\eta_-'))\nonumber\\
&&+2\bigg(\frac{\partial \ln \sigma_{ij} }{\partial\partial \theta_{i}}\bigg)
\bigg(\frac{\partial \ln \sigma_{ij} }{\partial\psi_{ij}}\bigg)+
\frac{\partial^2 \ln \sigma_{ij} }{\partial\psi_{ij}\partial \theta_{i}}\bigg|_{\rm corr},
\end{eqnarray}
\begin{eqnarray}
%%%%%%%%%%%%
\frac{\partial^2 \ln \sigma_{ij} }{\partial\theta_{i}^2}&=&\frac{\chi}{2}\bigg(\frac{\sigma_{ij}}
{\sigma_{ij}^0}\bigg)^2\bigg(\chi\cos(\theta_i-\theta_j)(\eta_+^2-\eta_-^2)-\nonumber\\
&&2\beta\cos(\theta_i-\psi_{ij})(\eta_++\eta_-)+
\frac{4\beta^2\sin^2(\theta_i-\psi_{ij})}{1-\chi^2\cos^2(\theta_i-\theta_j)}+\nonumber\\
&&4\beta\chi\sin(\theta_i-\psi_{ij})\sin(\theta_i-\theta_j)\nonumber\\
&&\bigg(\frac{\eta_+}{1+\chi\cos(\theta_j-\theta_i)}-\frac{\eta_-}{1-\chi\cos(\theta_j-\theta_i)}\bigg)+\nonumber\\
&&2\chi^2\sin(\theta_i-\theta_j)^2\nonumber\\
&&\bigg(\frac{\eta_+^2}{1+\chi\cos(\theta_j-\theta_i)}+
\frac{\eta_-^2}{1-\chi\cos(\theta_j-\theta_i)}\bigg)+\nonumber\\
&&2\bigg(\frac{\partial \ln \sigma_{ij} }{\partial \theta_{i}}\bigg)^2+
\frac{\partial^2 \ln \sigma_{ij} }{\partial\theta_{i}^2}\bigg|_{\rm corr},\\
%%%%%%%%%%%%
\frac{\partial^2 \ln \sigma_{ij} }{\partial\theta_{i}\partial \theta_{j}}&=&\frac{\chi}{2}\bigg(\frac{\sigma_{ij}}
{\sigma_{ij}^0}\bigg)^2(\cos(\theta_j-\theta_i)(\eta_+^2-\eta_-^2)-\nonumber\\
&&4\frac{\cos(\theta_j-\theta_i)\sin(\theta_i-\psi_{ij})
\sin(\theta_j-\psi_{ij})}{1-\chi^2\cos^2(\theta_i-\theta_j)}+\nonumber\\
&&2\beta\sin(\theta_i-\psi_{ij})\sin(\theta_j-\theta_i)\nonumber\\
&&\bigg(\frac{\eta_+}{1+\chi\cos(\theta_i-\theta_j)}-\frac{\eta_-}{1-\chi\cos(\theta_i-\theta_j)}\bigg)+\nonumber\\
&&2\beta^{-1}\sin(\theta_j-\psi_{ij})\sin(\theta_j-\theta_j)\nonumber\\
&&\bigg(\frac{\eta_+}{1-\chi\cos(\theta_i-\theta_j)}-\frac{\eta_-}{1+\chi\cos(\theta_i-\theta_j)}\bigg)-\nonumber\\
&&2\chi\sin^2(\theta_j-\theta_i)\nonumber\\
&&\bigg(\frac{\eta_+^2}{1-\chi\cos(\theta_i-\theta_j)}+\frac{\eta_-^2}{1+\chi\cos(\theta_i-\theta_j)}\bigg)-\nonumber\\
&&2\frac{\partial \ln \sigma_{ij} }{\partial\theta_{i}}
\frac{\partial \ln \sigma_{ij} }{\partial\theta_{j}}
-\frac{\partial^2 \ln \sigma_{ij} }{\partial\theta_{i}\partial \theta_{j}}\bigg|_{\rm corr},\\
%%%%%%%%%%%%
\eta_{\pm}&=&\frac{\beta\cos(\theta_i-\psi_{ij})\pm\beta^{-1}
\cos(\theta_j-\psi_{ij})}{1\pm \chi\cos(\theta_i-\theta_j)},\\
\eta_{\pm}'&=&\frac{\beta\sin(\theta_i-\psi_{ij})\pm\beta^{-1}
\sin(\theta_j-\psi_{ij})}{1\pm \chi\cos(\theta_i-\theta_j)},
\end{eqnarray}
and
%%%%%%%%%%%%
\begin{eqnarray}
\frac{\partial^2 \ln \sigma_{ij} }{\partial\psi_{ij}^2}\bigg|_{\rm  corr}&=&
\frac{\partial^2\sigma_{ij}(\lambda)}{\partial\lambda^2}
\bigg(\frac{\partial\lambda}{\partial\psi_{ij}}\bigg)^2,\\
%%%%%%%%%%%%
\frac{\partial^2 \ln \sigma_{ij} }{\partial\psi_{ij}\partial  \theta_{i}}\bigg|_{\rm
corr}&=&
\frac{\partial^2\sigma_{ij}(\lambda)}{\partial\lambda^2}
\frac{\partial\lambda}{\partial\psi_{ij}}\frac{\partial\lambda} {\partial\theta_{i}},\\
%%%%%%%%%%%%
\frac{\partial^2 \ln \sigma_{ij} }{\partial\theta_{i}^2}\bigg|_{\rm  corr}&=&
\frac{\partial^2\sigma_{ij}(\lambda)}{\partial\lambda^2}
\bigg(\frac{\partial\lambda}{\partial\theta_{i}}\bigg)^2,\\
%%%%%%%%%%%%
\frac{\partial^2 \ln \sigma_{ij} }{\partial\theta_{i}\partial  \theta_{j}}\bigg|_{\rm
corr}&=&
\frac{\partial^2\sigma_{ij}(\lambda)}{\partial\lambda^2}
\frac{\partial\lambda}{\partial\theta_{i}}\frac{\partial\lambda} {\partial\theta_{j}}.
%%%%%%%%%%%%
\end{eqnarray}


\begin{thebibliography}{99}
\bibitem{rattlers}
For systems with `rattler' particles that possess fewer than $d+1$ contacts, 
$N_{\rm dof} = N' d$ ($N'd -d$) for fixed (periodic) boundary conditions, 
where $N'=N-N_r$ and $N_r$ is the number of rattler particles.
\bibitem{foot_md} The CG gradient energy minimization technique we
implement relies on numerous evaluations of the total potential energy
to identify local minima.  However, when we integrate Newton's
equations of motion with damping forces proportional to particle
velocities (Eqs.~\ref{newton} and ~\ref{angle}), energy minimization
and the minimization stopping criteria are based on the evaluation of
forces and torques, which allows increased accuracy compared to the CG
technique.
\bibitem{foot_contact}
Since we employ overdamped energy minimization dynamics, rattler particles
can contact non-rattler particles with interparticle separation 
$r_{ij} = \sigma_{ij}$.
\end{thebibliography}
\end{document}